\documentclass[a4paper,fleqn,usenatbib]{mnras}

\usepackage{newtxtext,newtxmath}

\usepackage[T1]{fontenc}
\usepackage{ae,aecompl}

\usepackage{graphicx}	
\usepackage{amsmath}	
\usepackage{amssymb}	

\title[Rapid Optical Variations in V404 Cyg]{Rapid Optical Variations Correlated with X-rays in the 2015 Second Outburst of V404 Cygni (GS 2023$+$338)}

\author[M.~Kimura et al.]{Mariko Kimura,$^{1}$
\thanks{E-mail: mkimura@kusastro.kyoto-u.ac.jp}
Taichi Kato,$^{1}$
Keisuke Isogai,$^{1}$
Hyungsuk Tak,$^{2}$
Megumi Shidatsu,$^{3}$
\newauthor
Hiroshi Itoh,$^{4}$
Tam\'{a}s Tordai,$^{5}$
Kiyoshi Kasai,$^{6}$
William Goff,$^{7}$
Seiichiro Kiyota,$^{8}$
\newauthor
Roger D.~Pickard,$^{9,10}$
Katsura Matsumoto,$^{11}$
Naoto Kojiguchi,$^{11}$
Yuki Sugiura,$^{11}$
\newauthor
Eiji Yamada,$^{11}$
Taiki Tatsumi,$^{11}$
Atsushi Miyashita,$^{12}$
Pavol A.~Dubovsky,$^{13}$
\newauthor
Igor Kudzej,$^{13}$
Enrique de Miguel,$^{14,15}$
William L.~Stein,$^{16}$
Yutaka Maeda,$^{17}$
\newauthor
Elena P.~Pavlenko,$^{18}$
Aleksei A.~Sosnovskij,$^{18}$
Julia V.~Babina,$^{18}$
Lewis M.~Cook$^{19}$
\newauthor
and Daisaku Nogami$^{1}$
\\
$^{1}$Department of Astronomy, Graduate School of Science, Kyoto University, 
Oiwakecho, Kitashirakawa, Sakyo-ku, Kyoto 606-8502, Japan\\
$^{2}$Statistical and Applied Mathematical Sciences Institute, Durham, NC, USA\\
$^{3}$MAXI team, RIKEN, 2-1 Hirosawa, Wako, Saitama 351-0198, Japan\\
$^{4}$Variable Star Observers League in Japan (VSOLJ), 1001-105 Nishiterakata, 
Hachioji, Tokyo 192-0153, Japan\\
$^{5}$Polaris Observatory, Hungarian Astronomical Association, Laborc utca 2/c, 
1037 Budapest, Hungary\\
$^{6}$Baselstrasse 133D, CH-4132 Muttenz, Switzerland\\
$^{7}$American Association of Variable Star Observers (AAVSO), 13508 Monitor 
Lane, Sutter Creek, California 95685, USA\\
$^{8}$Variable Star Observers League in Japan (VSOLJ), 7-1 Kitahatsutomi, Kamagaya, 
Chiba 273-0126, Japan\\
$^{9}$The British Astronomical Association, Variable Star Section (BAA VSS), 
Burlington House, Piccadilly, London W1J 0DU, UK\\
$^{10}$3 The Birches, Shobdon, Leominster, Herefordshire HR6 9NG, UK\\
$^{11}$Osaka Kyoiku University, 4-698-1 Asahigaoka, Kashiwara, Osaka 582-8582, 
Japan\\
$^{12}$Seikei Meteorological Observatory, Seikei High School, Kichijoji-kitamachi 
3-10-13, Musashino, Tokyo 180-8633, Japan\\
$^{13}$Vihorlat Observatory, Mierova 4, Humenne, Slovakia\\
$^{14}$Departamento de F\'{i}sica Aplicada, Facultadde Ciencias Experimentales, 
Universidad de Huelva, 21071 Huelva, Spain\\
$^{15}$Center for Backyard Astrophysics, Observatorio del CIECEM, Parque Dunar, 
Matalasca\~{n}as, 21760 Almonte, Huelva, Spain\\
$^{16}$6025 Calle Paraiso, Las Cruces, New Mexico 88012, USA\\
$^{17}$12-2 Kaminishiyama-machi, Nagasaki, Nagasaki 850-0006, Japan\\
$^{18}$Crimean Astrophysical Observatory, 298409 Nauchny, Crimea\\
$^{19}$Center for Backyard Astrophysics (Concord), 1730 Helix Court, Concord, 
California 94518, USA\\
}

\date{Accepted XXX. Received YYY; in original form ZZZ}

\pubyear{2016}

\begin{document}
\label{firstpage}
\pagerange{\pageref{firstpage}--\pageref{lastpage}}
\maketitle

\begin{abstract}
\indent
   We present optical multi-colour photometry of V404 Cyg 
during the outburst from December, 2015 to January, 2016 
together with the simultaneous X-ray data.  
This outburst occurred less than 6 months after the previous 
outburst in June--July, 2015.  
These two outbursts in 2015 were of a slow rise and rapid 
decay-type and showed large-amplitude ($\sim$2 mag) and 
short-term ($\sim$10 min--3 hours) optical variations even 
at low luminosity (0.01--0.1$L_{\rm Edd}$).  
We found correlated optical and X-ray variations in two 
$\sim$1 hour time intervals and obtained a Bayesian estimate 
of an X-ray delay against the optical emission, which is 
$\sim$30--50 s, during those two intervals.  
In addition, the relationship between the optical and X-ray 
luminosity was $L_{\rm opt} \propto L_{\rm X}^{0.25-0.29}$ 
at that time.  
These features cannot be easily explained by the conventional 
picture of transient black-hole binaries, such as canonical 
disc reprocessing and synchrotron emission related to a jet.  
We suggest that the disc was truncated during those intervals 
and that the X-ray delays represent the required time for 
propagation of mass accretion flow to the inner optically-thin 
region with a speed comparable to the free-fall velocity.
\end{abstract}




\section{Introduction}

Transient low-mass X-ray binaries (LMXBs) are composed of 
a central neutron star or black hole and a low-mass companion 
star with an accretion disc around the central object.  
They show sporadic outbursts lasting for dozens of days up 
to several years in mainly X-rays and other wavelengths 
\citep{tan96XNreview}.  
The outbursts are considered to be caused 
by thermal-viscous instability over the accretion discs 
\citep[see chapter 5 of][]{kat08BHaccretion,las01DIDNXT,dub01XNmodel}.  
During their outbursts, reprocessing of 
X-ray irradiation in the outer cool discs has long been 
thought to dominate optical flux \citep{sha73BHbinary}.   
On the other hand, some LMXBs show rapid optical variations 
having timescales between milliseconds and minutes in 
quiescence and the low/hard state 
\citep[e.g.,][]{hyn04v404cyg,mot82gx339,uem02v4641sgrletter}.  
The origin of optical emission of these short-term variations 
is still unclear.  

   V404 Cyg is a member of these transient LMXBs.  This system 
hosts a 9${\rm M}_{\odot}$ black hole \citep{kha10v404cygcenx4} 
and a 0.7${\rm M}_{\odot}$ companion star of spectral type 
K0($\pm$1) III-V 
\citep{sha93v404cyg,wag92v404cyg,cas93v404cyg,hyn09v404cyg}.  
It is located at a distance of 2.4 kpc \citep{mil09v404cygdistance}.  
   It was originally discovered as a nova in 1938 and its 1989 
outburst was detected as an X-ray transient by the {\it GINGA} 
satellite \citep{mak89v404cygiauc4782}.  
At that time, the optical counterpart was subsequently identified 
with the 1938 nova \citep{wag89v404cygiauc4783}.  
During the 1989 outburst, short-term X-ray variability was 
observed \citep[e.g.,][]{zyc99v404cyg}.  
On June 15th in 2015, it underwent a short outburst after 26 
years of quiescence \citep{bar15v404cyggcn17929} and showed 
large flares in radio \citep{moo15atel7658,tet15v404cygatel7708}, 
infrared \citep{tan16v404cyg}, optical 
\citep{rod15v404cyg,mar16v404cyg,kim16v404cyg,gan16v404cyg}, 
X-ray \citep{kin15v404cyg,nat15v404cyg,neg15atel7646,seg15atel7755,wal17v404cyg,rad16v404cyg,jou17v404cyg,hup17v404cyg} 
and gamma-ray wavelengths 
\citep{roq16v404cyg,sie16v404cyg,jen16v404cyg,loh16v404cyg,pia17v404cyg}.  

   This system exhibited violent optical variations with 
regular patterns for some intervals during 
the 2015 June outburst, when most of the optical flux was 
likely to be produced by the reprocessing of X-ray irradiation 
in the outer disc \citep{kim16v404cyg}.  
A lack of optical and near-infrared polarisation was 
reported by \citet{tan16v404cyg} in two 
periods during the outburst.  
There was, however, evidence of a strong contribution of 
synchrotron emission related to jet ejections in 
some other epochs 
\citep{rod15v404cyg,mar16v404cyg,sha16v404cyg,lip16v404cyg,ber16cenX4v404cyg}.  
\citet{gan16v404cyg} found sub-second optical flaring events 
and proposed that not only X-ray reprocessing but also 
non-thermal emission contributed to them.  
Thus the origin of optical emission in the outburst is still 
under debate.  

   At 05:19:52 UT on December 23rd, 2015, the {\it Swift} Burst 
Alert Telescope (BAT) initially detected that the X-ray flux 
increased above the detection limit \citep{bar15bGCNv404cyg}.  
Just after the BAT detection, MASTER-Amur began 
observing 
this object on December 23.385 UT in optical wavelengths 
\citep{lip15bv404cyg}.  
\citet{mun16v404cyg2} reported evidence of a strong wind 
with their optical spectroscopy and the multi-wavelength 
variability, which were very similar to those in the June 
outburst.  
In this paper, we report on our optical photometry of the 
December outburst in V404 Cyg and study their correlation 
with the simultaneous X-ray data of {\it INTEGRAL} Imager 
on Board the Integral Satellite (IBIS)/CdTe array (ISGRI) 
monitoring.

\section{Observation and Analysis}

\subsection{Optical Observations}

   Time-resolved CCD photometry was carried out by the Variable 
Star Network (VSNET) collaboration team \citep{VSNET} at 17 sites 
(Table S1) in the 2015 December outburst in V404 Cyg.  
Table S2 shows the log of our photometric observations in 
the $V$, $R_{\rm C}$ and $I_{\rm C}$ bands and with a clear 
filter.  The exposure times were 15--540 s.  
We also used the data downloaded from the American Association of 
Variable Star Observers (AAVSO) 
archive\footnote{<http://www.aavso.org/data/download/>}.  
All of the observation times were converted to barycentric 
Julian date (BJD).  The comparison stars are listed in Table S3.  
The constancy of the comparison stars was checked by nearby 
stars in the same images.  
The data reduction and the calibration of the comparison stars 
were performed by each observer.  The magnitude of each comparison 
star was measured by A.~Henden from the AAVSO Variable Star 
Database\footnote{<http://www.aavso.org/vsp>}.  
The tables are displayed in the supplements to this paper.

\subsection{X-ray Analysis}

   We extracted X-ray light curves with time bin sizes of 
1 s and 5 s in the 25--60 keV energy band to use them for timing 
analyses (Sec.~3.2 and 3.3) from the archived data of the 
{\it INTEGRAL} IBIS/ISGRI monitoring set.  We employed the 
latest version of the standard data analysis software 
\texttt{Off line Scientific Analysis (OSA) 
v.10.2}\footnote{<http://www.isdc.unige.ch/integral/analysis\#Software>} 
for pipeline processing.  
The publicly available pointing observations between MJD 
57387.65--57387.80 are composed of 4 science windows (SCWs).  
Each of the SCWs has a typical good time of $\sim$3 ks in 
duration.  
An image in the 25--60 keV energy band was generated from 
IBIS-ISGRI data by using an input catalogue, 
gnrl\_refr\_cat\_0039.fits.  
Background maps provided by the ISGRI team were used for 
background correction.  We put gnrl\_refr\_cat\_0039.fits 
[ISGRI\_FLAG2==5 \&\& ISGR\_FLUX\_1$>$100] (the default 
parameter) in the parameter ``brSrcDOL'' in the IBIS Graphical 
User Interface.  
The spectra and 1-s and 5-s binned light curves were extracted 
with the tool \texttt{ii\_light} and a catalogue composed of 
the strong sources in the field of view (FOV); AX J1949.8$+$2534, 
Cygnus X$-$1, Cygnus X$-$3 and Ginga 2023$+$338 (V404 Cyg).  
   We derived the X-ray light curves in flux scales 
using a conversion parameter of 1 [count s$^{-1}$] to be 
5.632$\times 10^{-11}$ [ergs cm$^{-2}$ s$^{-1}$] with the 
HEAsoft package assuming a Crab-like spectrum.  
   Moreover, we obtained lists of photons in the SCWs with the 
tool \texttt{ii\_pif} to use them for power spectral analyses 
(Sec.~3.3).  All of the observation times were converted to BJD.  

   To estimate the bolometric correction factor 
$L_{\rm bol} / L_{25-60 \rm keV}$\footnote{Here, $L_{\rm bol}$ is 
the bolometric luminosity.}, we analysed multi-wavelength 
spectral energy distributions (SEDs) in two intervals 
(during BJD 2457382.85--2457382.86 and BJD 2457386.58--2457386.60), 
when the source was fully simultaneously observed in X-rays 
({\it Swift}/X-ray Telescope (XRT) data, obsID: 00031403127 and 
00031403131, respectively) and optical bands, assuming that they 
can be reproduced with the same irradiated disc model employed 
by \citet{kim16v404cyg} for the 2015 June--July outburst.  
The estimated bolometric correction factors in these 
two intervals were 9.1 (ID 00031403127) and 10 (ID 00031403131), 
respectively.  
These values were almost the same as that of the estimated 
correction factor (9.97) in the previous outburst \citep[see ED 
Fig.~6(a) in][]{kim16v404cyg}.  We then adopted the same 
correction factor, 9.97, to estimate $L_{\rm bol}$ from the 
{\it INTEGRAL} IBIS/ISGRI X-ray 25--60 keV 
light curves in the 2015 December outburst.

\begin{figure*}
\begin{center}
\includegraphics[width=17cm]{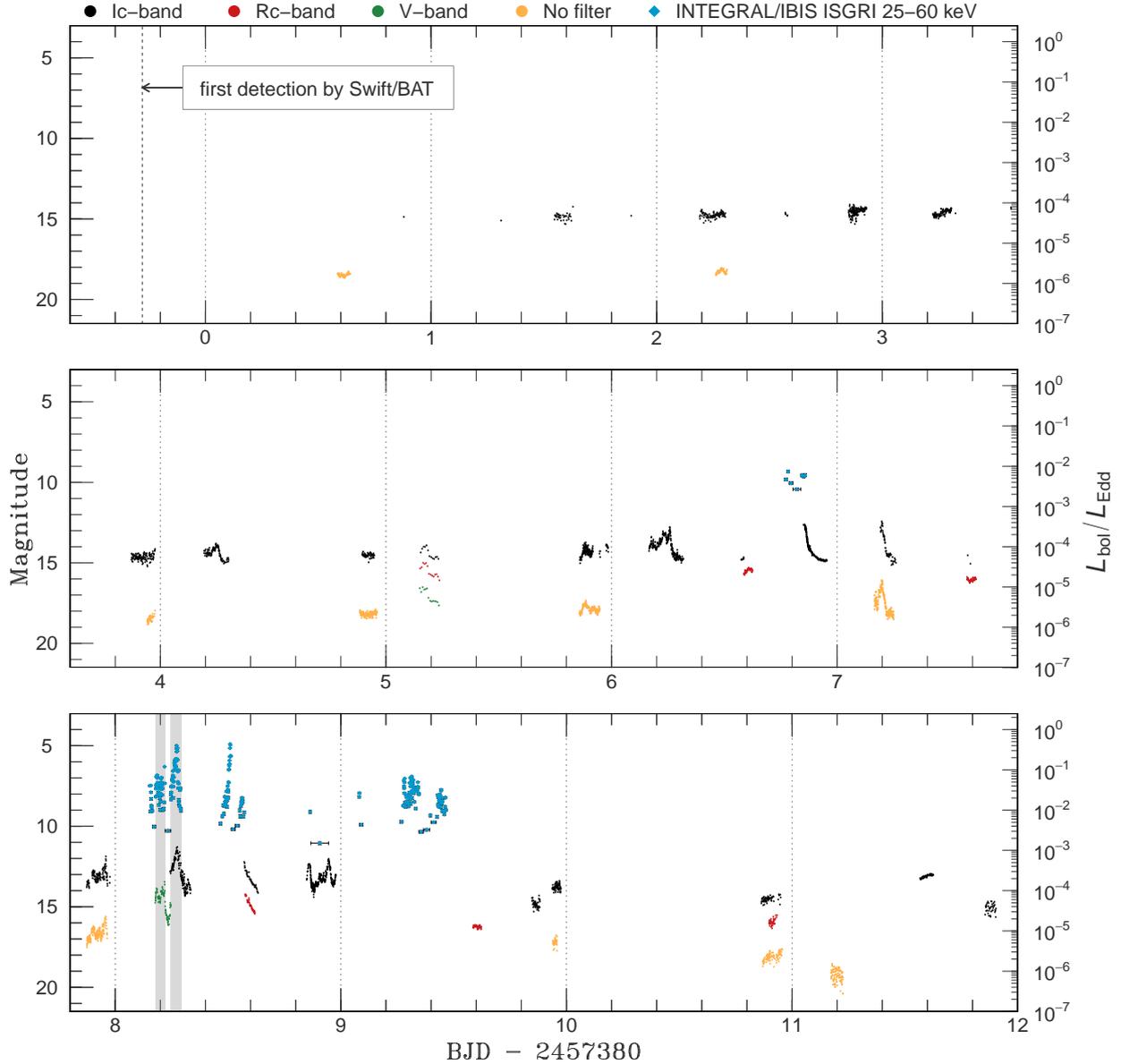}
\end{center}
\caption{Overall light curves in the optical $I_{\rm C}$, 
$R_{\rm C}$, $V$ bands and with no filter and in the X-ray 
25--60 keV energy band of {\it INTEGRAL} IBIS/ISGRI monitoring 
during the 2015 December outburst in V404 Cyg.  
For clarity, the plotted magnitude for the unfiltered data is 
fainter by 2 mag than measured.  The horizontal axis represents 
days from BJD 2457380.  
Here, $L_{\rm Edd}$ is equal to $1.35 \times 10^{39}$ [erg/s].  
For visibility, only data points having horizontal error bars 
less than 1 hour are plotted.  
The grey shadings represent the overlapped optical and X-ray 
observational periods.  }
\label{multi}
\end{figure*}

\section{Results}

\subsection{Rapid Optical Variations}

   We detected large-amplitude and short-term optical 
variations with amplitudes ranging from 0.4 to 2.5 mag 
on timescales of $\sim$10 min--3 hours during the 2015 
December outburst in V404 Cyg.  
The bolometric luminosity derived from the X-ray flux with 
the correction factor (see Sec.~2.2) was low.  
The overall optical light curves of the outburst in the 
$I_{\rm C}$, $R_{\rm C}$, $V$ and no-filter 
bands, and the X-ray 25--60 keV light curves of 
{\it INTEGRAL} IBIS/ISGRI monitoring with time bin size of 
64 s downloaded from the archive 
data\footnote{<http://www.isdc.unige.ch/integral/analysis\#QLAsources>} 
\citep{kul16atel8512} are displayed in Figure 
\ref{multi}.\footnote{The X-ray light curves are 
adaptively rebinned to achieve signal-to-noise ratio (S/N) 
$>$ 8 by the {\it INTEGRAL} team.}
Hereafter, we choose BJD 2457380 as the time reference and 
report the time from that day.  
Sudden dips in brightness were detected for several 
time intervals (during the day 4.18--4.31 for example).  
   The variations with amplitudes of $\gtrsim$2 mag were 
observed only when the nightly average magnitude was brighter 
than $\sim$14.2 mag in the $I_{\rm C}$ band, in the middle 
term of the outburst.  
The maximum magnitude in the brightest interval 
during the December outburst (the day 8.24--8.34) was 11.3 mag 
in the $I_{\rm C}$ band and the maximum bolometric luminosity in 
that interval was 4.37$\times 10^{-1} L_{\rm Edd}$.  
   The average brightness gradually increased during the day 
0--7.5, was constant during the day 7.5--8.8 and rapidly decreased 
during the day 8.8--11.3.  A small rebrightening was observed 
soon after the rapid decay (see the $I_{\rm C}$-band light 
curves in Figure \ref{multi}).  

\begin{figure}
\begin{center}
\includegraphics[width=9cm]{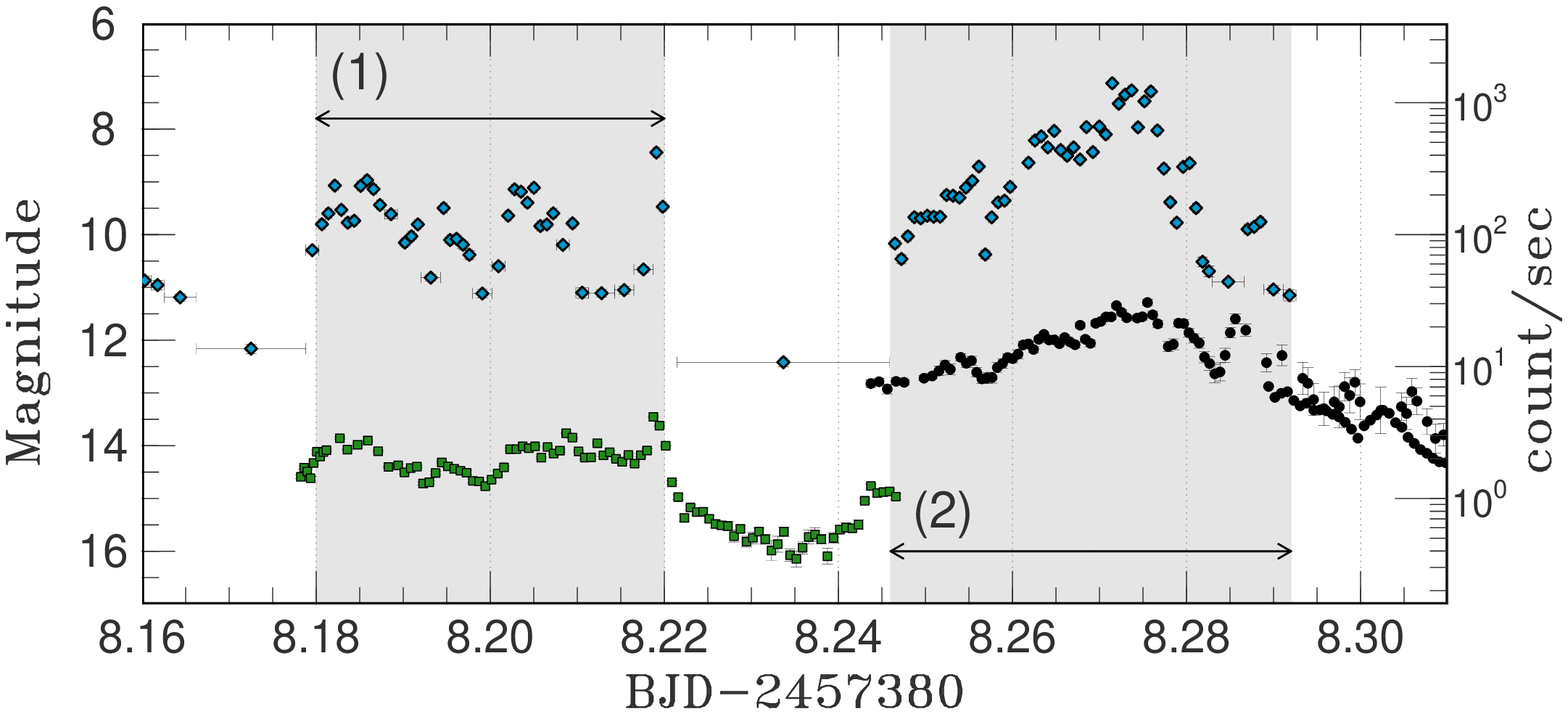}
\end{center}
\caption{Simultaneous optical and X-ray light curves during (1) the day 8.18--8.22 and (2) the day 8.246--8.292.  Each of the intervals is $\sim$1 hour long.  The blue rhombuses, green squares and black circles represent the X-ray 25--60 keV, optical $V$-band and optical $I_{\rm C}$-band light curves.  The optical flares are broader than the X-ray ones.}
\label{datacorr}
\end{figure}

\begin{figure*}
\begin{minipage}{.49\textwidth}
\label{relation1}
\begin{center}
\includegraphics[width=8cm]{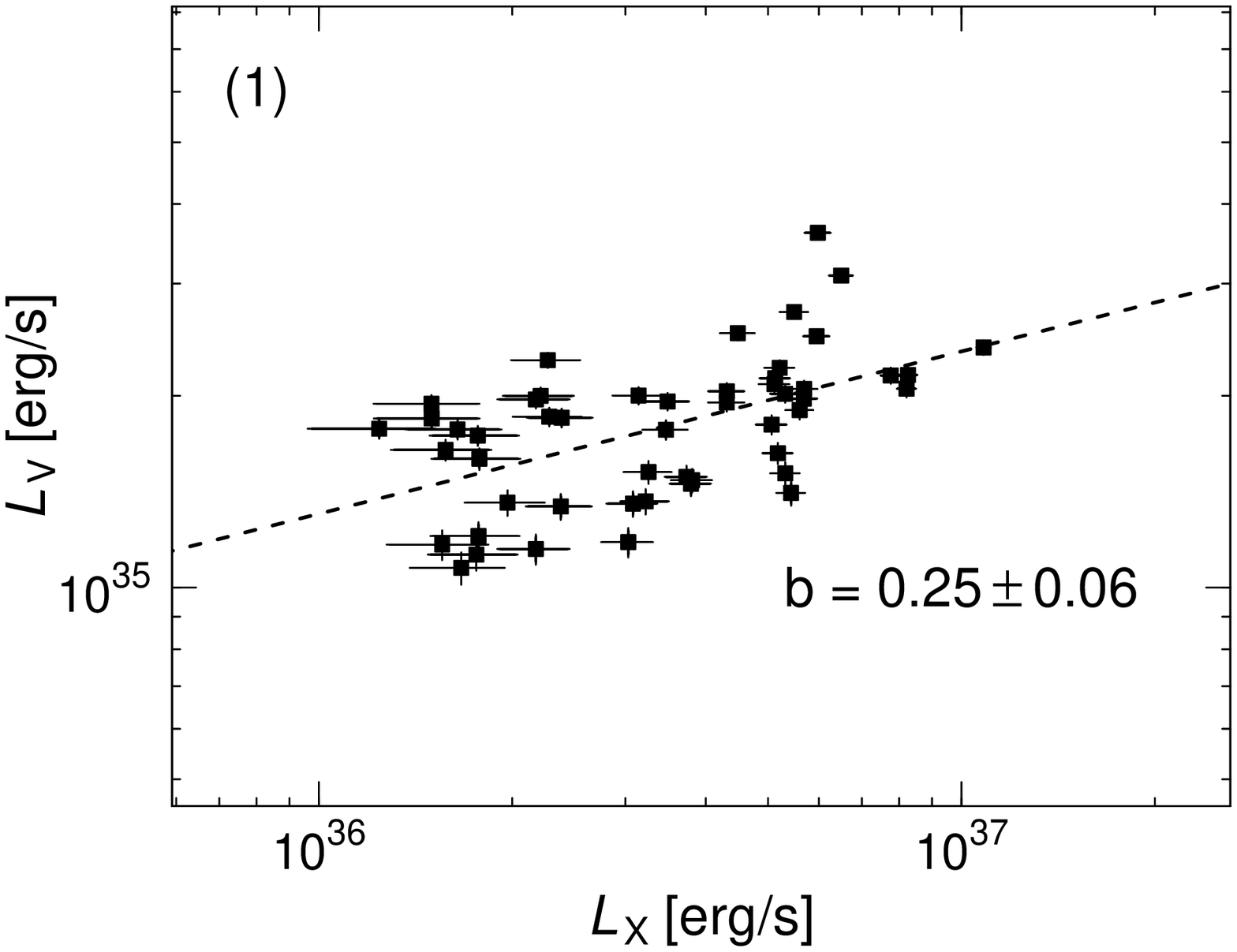}
\end{center}
\end{minipage}
\begin{minipage}{.49\textwidth}
\label{relation2}
\begin{center}
\includegraphics[width=8cm]{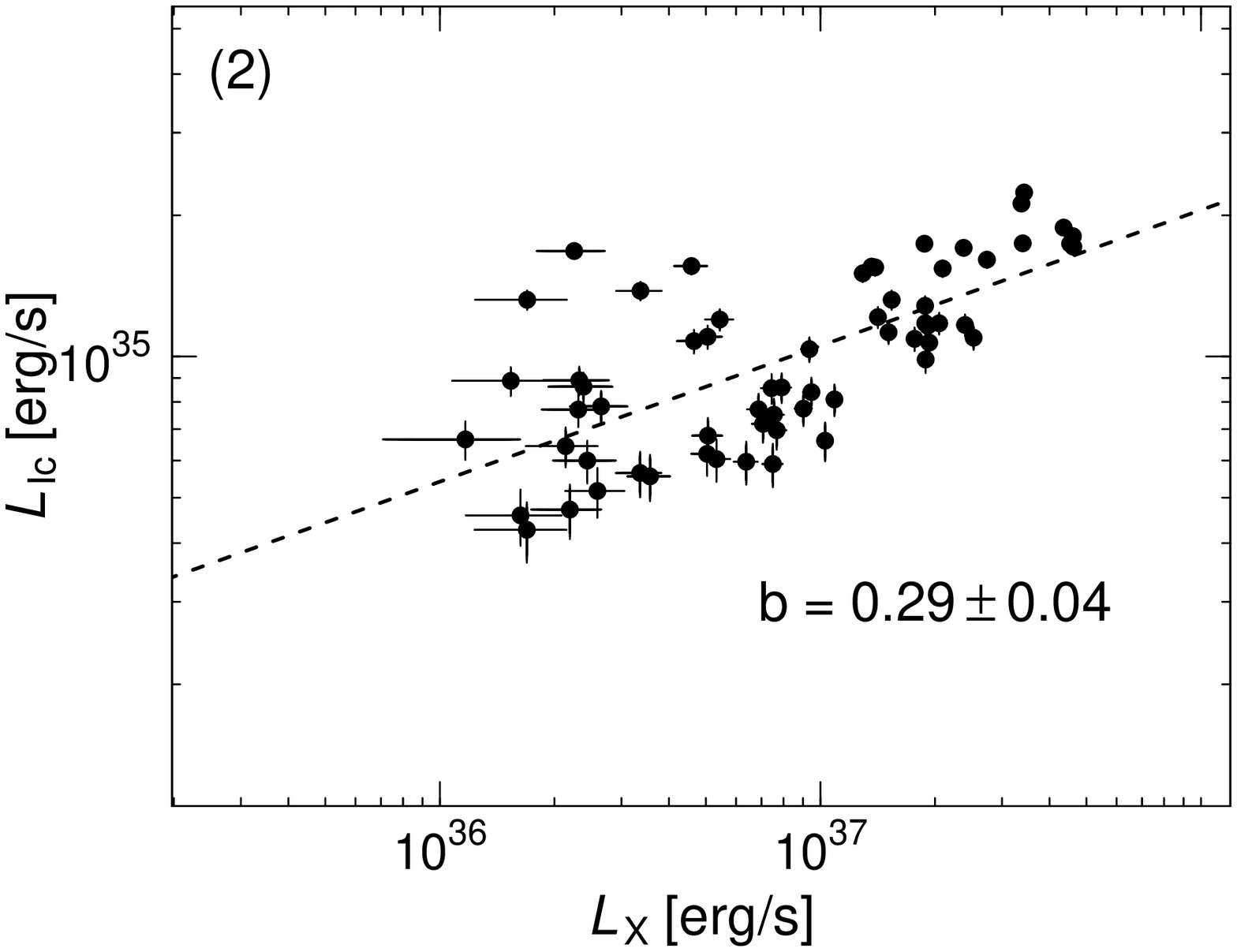}
\end{center}
\end{minipage}
\caption{Optical and X-ray correlations during interval 
(1) on the day 8.18--8.22 (the left panel) and interval 
(2) on the day 8.246--8.292 (the right panel).  The filled 
squares and circles represent the optical $V$-band and 
$I_{\rm C}$-band luminosity.  
The horizontal axes exhibit X-ray luminosity in the 25--60 keV 
energy band.  The dashed lines represent the estimated power law 
regression formulae for the relations between the optical and 
X-ray luminosity.  The values of the power law index ($b$) are 
also reported.  }
\label{relation}
\end{figure*}

\subsection{Optical and X-ray Correlation}

   We found the simultaneous X-ray data with our optical data 
for two intervals (1) during the day 8.18--8.22 and (2) during 
the day 8.246--8.292 where stochastic variations were observed 
(see the grey shadings in Figure \ref{multi}).\footnote{
Althogh some of {\it INTEGRAL} ToO observations were performed 
during some other intervals including the periods in which 
high-amplitude optical variations were observed (e.g., the day 
6.16--6.32, the day 7.16--7.35 and the day 8.84--8.96), the 
luminosity was very low at that time and it was difficult to 
judge whether the optical and X-ray variability was correlated 
or not in those intervals due to the low S/N of the X-ray data.}
The enlarged view of the intervals is displayed 
in Figure \ref{datacorr}.  
   To examine the correlations between the optical and X-ray 
luminosity, we performed Spearman's rank tests on them in 
logarithmic scales and power law regression assuming that 
they follow $y=ax^{b}$.  Here, $x$ and $y$ denote the X-ray 
and optical ($V$ or $I_{\rm C}$ band) luminosity in linear 
scales, respectively.  In these analyses, the completely 
simultaneous X-ray light curves to the optical ones were 
obtained by binning the {\it INTEGRAL} IBIS/ISGRI 25--60 keV 
light curves having 1-s time bin size\footnote{The 1-s binned 
X-ray light curves were derived with the tool \texttt{ii\_light} 
described in Sec.~2.2.} with the exposure times 
of the optical observations (30 s for interval 
(1) and 50 s or 60 s for interval (2)) and 
calculating weighted averaged flux in each bin.  The optical 
data are corrected for interstellar extinction/absorption 
by assuming $A_{\rm V}$ = 4 \citep{cas93v404cyg,car89extinction}.  
The results are summarised in Table \ref{fitting}.  Both 
analyses have confirmed the correlated variations.  
Optical vs.~X-ray luminosity during the two 
time intervals is displayed in Figure \ref{relation} with the 
regression equations.  

\begin{table}
	\centering
	\caption{Results of Spearman's rank tests and power 
	law regression for the model of form, $y=ax^{b}$, 
	in time intervals 
	(1) on the day 8.18--8.22 and (2) on the day 8.246--8.292.}
	\label{fitting}
	\begin{tabular}{ccccc}
		\hline
		Intervals & $a^{*}$ & $b^{*}$ & $\rho$$^{\dagger}$ & $p$-value$^{\ddagger}$\\
		\hline
(1) & $10^{(25.9 \pm 2.076)}$ & 0.255 ($\pm$0.0568) & 0.60 & $4.546 \times 10^{-6}$\\
(2) & $10^{(24.3 \pm 1.431)}$ & 0.290 ($\pm$0.0388) & 0.71 & $4.573 \times 10^{-11}$\\
\hline
\multicolumn{5}{l}{\parbox{240pt}{$^{*}$The values of degree 
of freedom (d.o.f) of the regression in time intervals 
(1) and (2) are 49 and 61, respectively.  As a result of 
$t$-tests of the regression coefficients, $t$-values of 
$a$ and $b$ were 12.495 ($p < 2 \times 10^{-16}$) and 4.487 
($p = 4.37 \times 10^{-5}$) for interval 
(1) and 16.988 ($p < 2 \times 10^{-16}$) and 7.472 ($p = 3.57 
\times 10^{-10}$) for interval (2), 
respectively.  These tests prove the estimated values of 
$a$ and $b$ to be statistically significant at the 1\% level.  
In addition, the values of $F$-tests were 20.84 ($p = 4.371 
\times 10^{-5}$) for interval (1) and 55.83 
($p = 3.567 \times 10^{-10}$) for interval 
(2), respectively, and hence, the assumption that $a = b = 0$ 
is abandoned.  }}\\
\multicolumn{5}{l}{$^{\dagger}$Spearman's correlation 
coefficients.}\\
\multicolumn{5}{l}{\parbox{240pt}{$^{\ddagger}$Null 
hypothesis probability of Spearman's correlation test.  
The null hypothesis is abandoned because the probability 
values are less than 0.01.  }}\\
\end{tabular}
\end{table}

\begin{figure}
\begin{center}
\includegraphics[width=9cm]{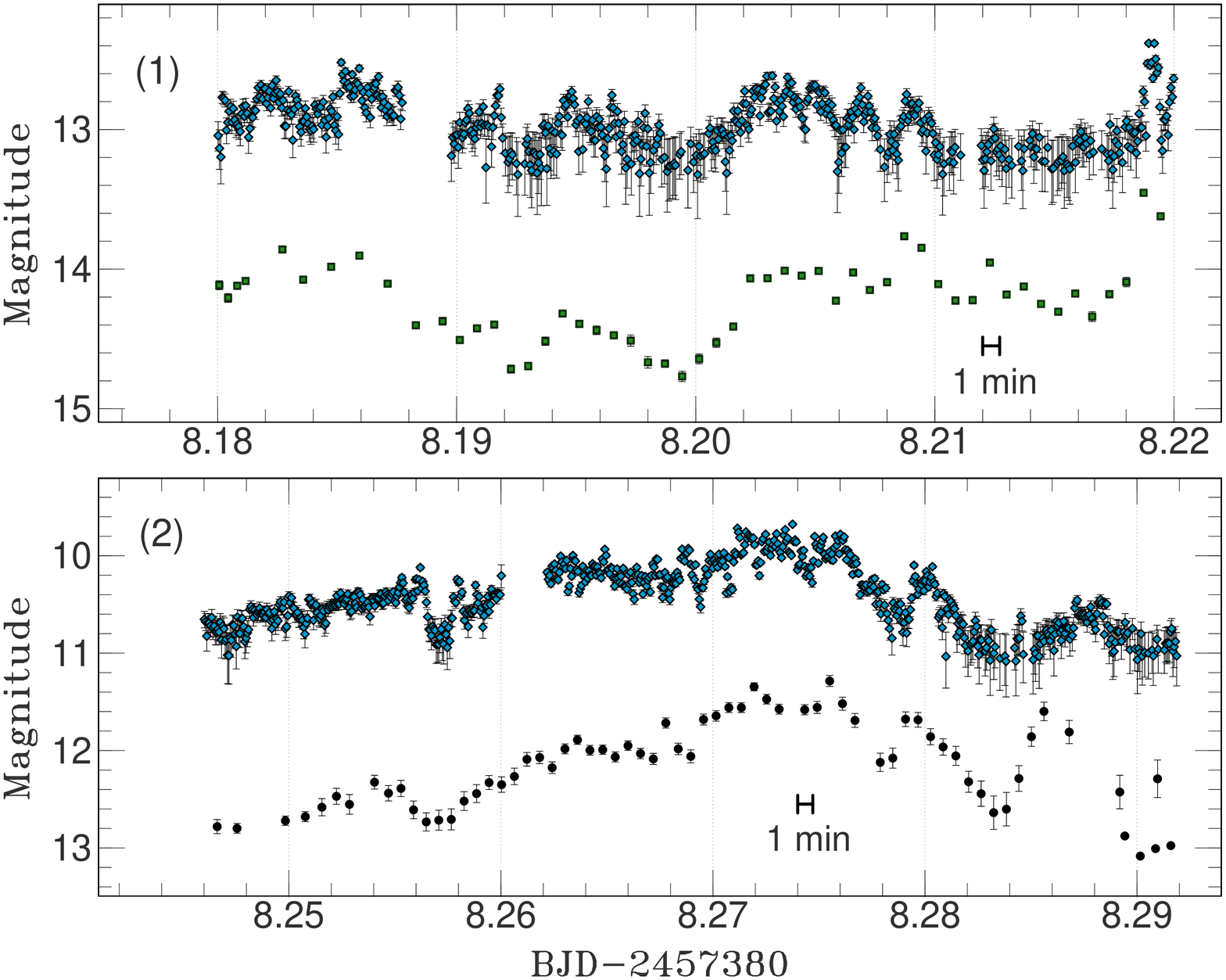}
\end{center}
\caption{X-ray and optical data sets used for 
the time delay estimates in interval (1) during the day 8.18--8.22 
and interval (2) during the day 8.246--8.292.  The X-ray light 
curves are rescaled by using the results of power 
law regression described in Sec.~3.2.  
The rhombuses, squares and circles represent the X-ray light curves 
in the 25--60 keV energy band with time bin size of 5 s after scaling, 
the optical $V$-band light curves and the optical $I_{\rm C}$-band 
light curves.  For visibility, the X-ray magnitudes 
are offset by 7.7 for interval (1) and 4.7 for interval (2), 
respectively.  }
\label{datafortimelag}
\end{figure}

\begin{figure*}
\begin{minipage}{.49\textwidth}
\label{td1}
\begin{center}
\includegraphics[width=8cm]{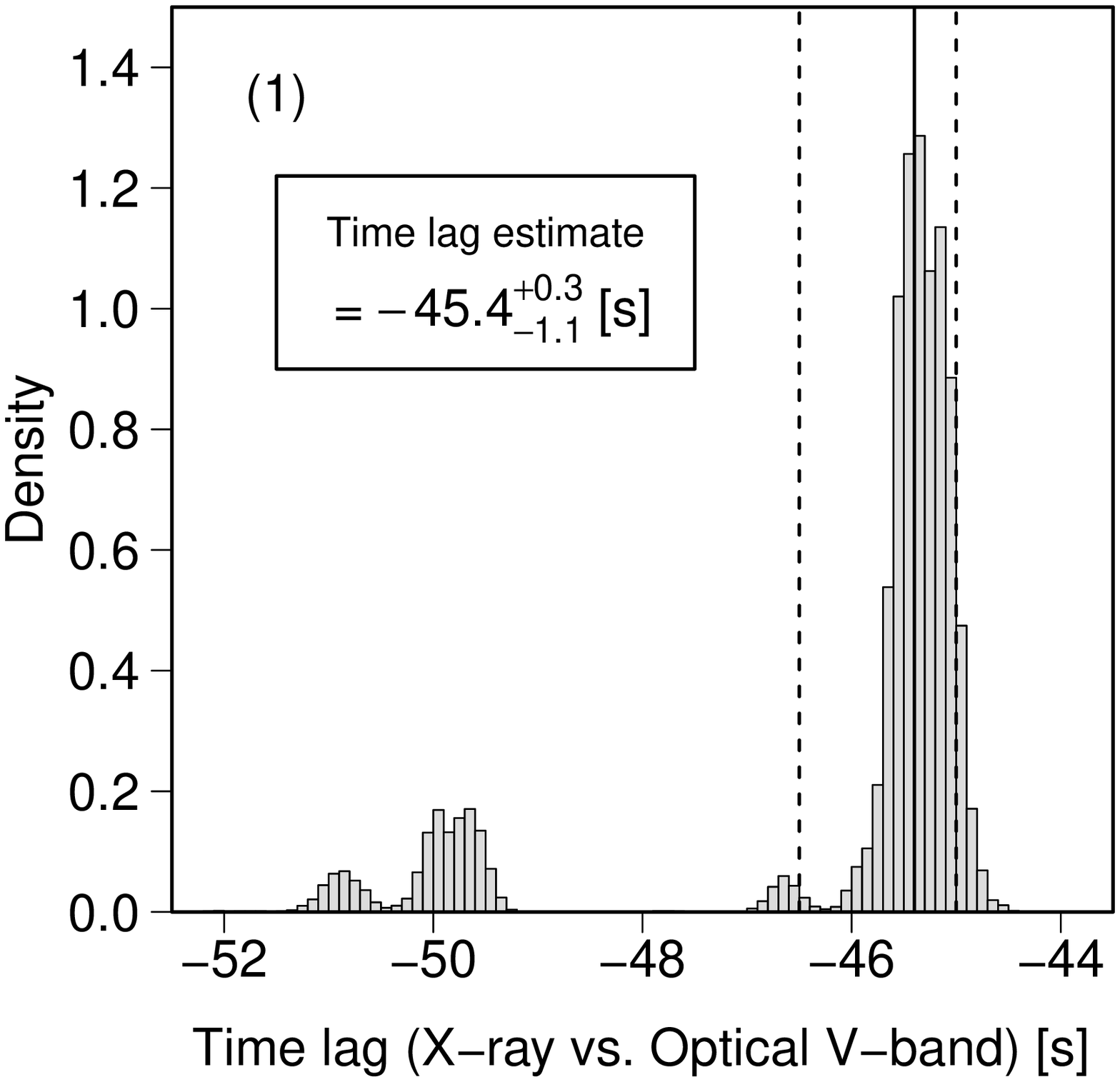}
\end{center}
\end{minipage}
\begin{minipage}{.49\textwidth}
\label{td2}
\begin{center}
\includegraphics[width=8cm]{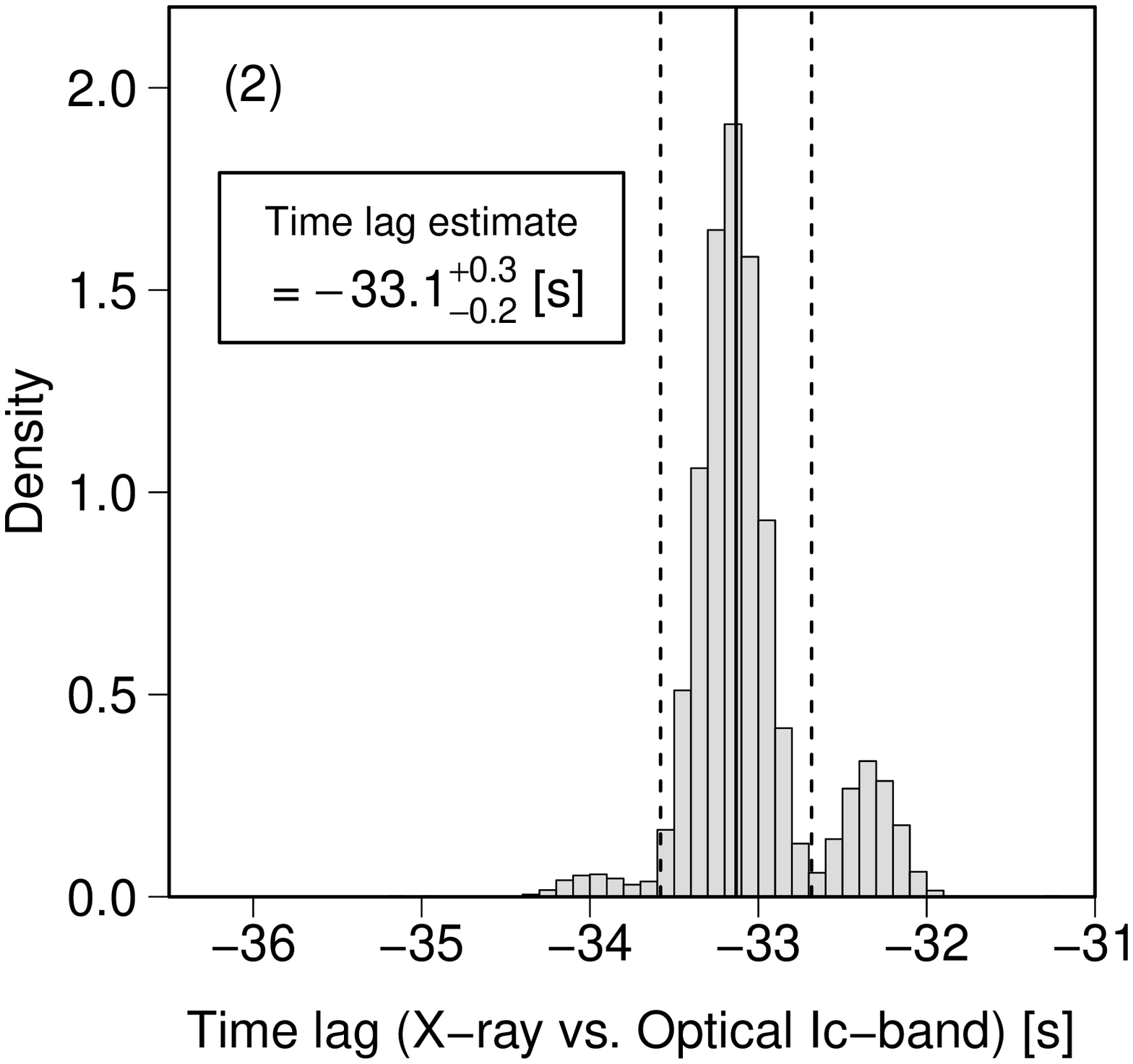}
\end{center}
\end{minipage}
\caption{Posterior distributions of the time delays of the 
optical variations against the X-ray ones for interval 
(1) on the day 8.18-8.22 (the left panel) and 
interval (2) on the day 8.246-8.292 (the right 
panel).  
The solid line indicates the posterior median of the time lag 
and the dashed lines represent the 68\% quantile-based interval.  
The time lag estimate shown in each figure is the posterior 
median with 68\% quantile-based interval.  There are invisibly 
small modes near $-$30.5 s and $-$25.8 s in the posterior 
distribution for interval (1) and near $-$48.1 s 
in that for interval (2), but we displayed 
only major modes.}
\label{timedelayfig}
\end{figure*}

\subsection{Bayesian Time Delay Analysis}

   We estimated time delays between our optical light curves 
and the {\it INTEGRAL} IBIS/ISGRI X-ray light curves with time 
bin size of 5 s in the 25--60 keV energy band for 
interval (1) on the day 8.18--8.22 and 
interval (2) on the day 8.246--8.292 using 
a Bayesian method that was originally proposed to estimate 
time delays between gravitationally lensed stochastic light 
curves \citep{tak16timedelay}.\footnote{The 5-s binned X-ray 
light curves were derived with the tool \texttt{ii\_light} 
described in Sec.~2.2.}  
This method assumes that the irregularly sampled light curves 
are generated by a latent continuous-time damped random walk 
(DRW) process \citep{kel09QuasarOptical} and that one of the 
latent light curves is a shifted version of the other by the 
time lag in the horizontal axis and by the magnitude offset 
in the vertical axis \citep{pel94timedelay}.  
The model also adopts heteroskedastic Gaussian measurement 
errors.  

   The DRW process is a stochastic process to describe a 
random walk with a tendency to move back towards a central 
location.  
This process is known to be appropriate for modelling 
accretion-type light variation such as the variability 
observed in active galactic nuclei (AGNs) because of its 
power-law type power spectral density 
\citep{kel09QuasarOptical,koz10QuasarVari,mac10DRW}.  
This process is also suitable for modelling another 
accretion-type light variation in a black-hole binary.  
This is because the power spectral densities (PSDs) of the 
X-ray light variations in V404 Cyg in the SCWs (162800020010, 
162800030010, 162800040010 and 162800050010) including 
intervals (1) and (2) are well expressed 
by a power-law ($P \propto f^{-\Gamma}$) with an index 
$\Gamma$ of 1.6$\pm$0.1, 1.5$\pm$0.1, 1.6$\pm$0.1 and 
1.0$\pm$0.2, respectively (see also Figure 
S1 in the supplements to this paper).\footnote{We employed 
powerspec software in the FTOOLS Xronos package from the 
lists of photons.  The values of ``dtnt'' and ``rebin'' 
parameters were 1 s and $-$1.8, respectively.  The Nyquist 
frequency of these observations was 0.5 Hz.  }

   Our data, however, do not completely meet the second model 
assumption (i.e., one of the latent light curves 
is a parallel-shifted version of the other).  This is because 
our optical light curves have smaller amplitudes 
than the X-ray ones and the origin 
of the optical light curves may be different from that of 
the X-ray ones unlike gravitationally lensed light curves 
as originally applied in \citet{tak16timedelay}.  
Thus we scaled the X-ray light curve to the optical one using 
the results of the power law regression (see Sec.~3.2 and Table 
\ref{fitting}) to meet the assumption before implementing 
the Bayesian method; we treated the scale change in the X-ray 
light curve as a known constant.  
The data sets for our analyses are displayed 
in Figure \ref{datafortimelag}.  X-ray emission seems to be 
delayed to the optical emission when we focus on 
the sharp peaks with small measurement errors (e.g., around the 
day 8.209 in the upper panel and around the day 8.279 in the 
lower panel).  
The following time lag estimations enable us to investigate 
the delay quantitatively.  

   We used an R package, \texttt{timedelay}, which we made 
available to the public at 
CRAN\footnote{<https://cran.r-project.org/package=timedelay>}, 
to implement the Bayesian model via a Markov chain Monte Carlo 
(MCMC) method; see Appendix A for details of the model, 
implementation, and model checking.  
Figure \ref{timedelayfig} exhibits the histogram of 300,000 
posterior samples of the time lag between the optical and 
X-ray light curves for interval (1) on the 
left panel and that for interval (2) on the 
right panel.  The estimation results are summarised in Table 
\ref{bayesian-result}; the posterior median of the time delay 
for interval (1) was $-$45.3$_{-1.1}^{+0.3}$ s 
and that for interval (2) was 
$-$33.1$_{-0.2}^{+0.3}$ s, i.e., the X-ray variations were 
delayed to the optical variations by 45.3$_{-1.1}^{+0.3}$ s 
for interval (1) and 33.1$_{-0.2}^{+0.3}$ s 
for interval (2).  
The Gelman-Rubin convergence diagnostic statistics 
\citep{gelmanrubin92} were 1.0004 and 1.0009 in 
intervals (1) and (2), respectively, close 
enough to unity.  
   To check the consistency between different estimation 
methods, we compared our estimates with the results 
of a locally normalized discrete correlation function 
\cite[LNDCF;][]{leh92timedelay}.  The LNDCF estimates 
averaged with time bin size equal to 25.92 s showed 
$-$35$_{-13}^{+13}$ s time lags in both intervals (1) and (2) 
(see also Figure S2 in the supplements 
to this paper).  
Both Bayesian and LNDCF methods result in consistent estimates, 
considering the large uncertainties of the LNDCF estimates.  

\begin{table}
	\centering
	\caption{Bayesian estimates of the time delays for 
	interval (1) during the day 8.18--8.22 
	and interval (2) during the day 
	8.246--8.292.  
	The 68\% interval indicates the quantile-based interval 
	and the 68\% HPD interval represents the highest posterior 
	density interval.}
	\label{bayesian-result}
	\begin{tabular}{cccc}
		\hline
		Intervals & Median$^{*}$ & 68\% Interval & 68\% HPD Interval\\
		\hline
(1) & $-$45.4 s & ($-$46.5 s, $-$45.1 s) & ($-$45.6 s, $-$45.0 s) \\
(2) & $-$33.1 s & ($-$33.3 s, $-$32.8 s) & ($-$33.4 s, $-$32.9 s) \\
\hline
\multicolumn{4}{l}{\parbox{220pt}{$^{*}$We report posterior 
medians because posterior means are not reliable indicators 
for the centre of a multi-modal distribution.  The posterior 
mode and median of time delays are identical up to three 
decimal places for interval (1).  
For interval (2), the posterior 
mode is $-$33.2 s.  }}\\
\end{tabular}
\end{table}

\section{Discussion}

\subsection{Similarities between the Two Outbursts in 2015}

   Large-amplitude and short-term optical variations at low 
luminosity, which have good correlations with simultaneous 
X-ray variability, were observed during the two outbursts 
in 2015 in V404 Cyg \citep[see also][]{kim16v404cyg}.  
This behaviour seems to be a common feature 
in every outburst of this system.  Actually, violent optical 
variations with amplitudes of $\sim$1 mag on timescales of 
days or minutes were observed also in the late stage of the 
1989 outburst \citep{wag91v404cyg}.  
The amplitudes and timescales of these optical variations 
and the occasional sudden dips in brightness during the 
December outburst were similar to those during the June 
outburst.  

   The overall trend of our optical light curves (a slow 
rise and rapid decay) in the December outburst was also similar 
to that in the June/July outburst as \citet{mun16v404cyg2} 
already pointed out.  
This trend is different from the most common type of 
outburst in transient LMXBs \citep{che97BHXN}.  
The slow rise could be the result of an inside-out outburst 
which is considered to arise more frequently 
than an outside-in outburst in these objects 
\citep[][for a review]{las01DIDNXT}.  
This possibility has already been suggested during the June 
outburst by estimating the disc radius at which an optical 
precursor was ignited \citep{ber16v404cyg}.

\subsection{Differences in the Short-term Variability between 
the Two Outbursts in 2015}

   Although the morphology of violent and rapid optical 
variations during the June and December outbursts resemble 
each other, the nature of these variations during the December 
outburst seems to be different from those during the June 
outburst.  
   This is because we found X-ray variations lagging optical 
ones by $\sim$30--50 s for the two intervals during the December 
outburst (Sec.~3.3), which had not been detected during the 
June outburst.  
   We consider whether two commonly known mechanisms of 
optical emission in X-ray binaries could explain the X-ray 
delays.  
They were expected to have been dominant at least some time intervals 
during the June outburst.  The first one is the reprocessing 
of X-ray irradiation from the inner disc 
\citep[e.g.,][]{van94visualLMXB} 
and the second one is cyclo-synchrotron emission from 
magnetic flares, which would be related to a jet 
\citep[e.g.,][]{mer00j1118,mar01j1118}.  
X-ray reprocessing will produce optical 
delays against X-rays on timescales of tens of seconds 
\citep[e.g.,][]{hyn98j1655echo}.  
On the other hand, cyclo-synchrotron radiation will induce 
very short time lags within 1 s between optical and X-ray 
variations, which are not consistent with the lags caused 
by X-ray reprocessing 
\citep[e.g.,][]{kan01j1118varcorrelation}.\footnote{There 
may be a possibility that the very short time lag is not found 
because the observational data used in this work do not allow 
investigating millisecond-scale timing properties.  }
We, therefore, conclude that these processes would fail to 
explain the observed X-ray delays in the December outburst.  
   Even if we observed the expanding jet ejecta, the expected 
time lag was a $\gtrsim$10 min optical delay 
\citep{van66RadioModel,mir98grs1915} as discussed also 
in the June outburst \citep{rod15v404cyg,mar16v404cyg}.  
It is quite different from the X-ray delays that we estimated.  

   There are some other models including synchrotron radiation, 
which have been developed to explain recently detected 
anti-correlated cross-correlation function (CCF) 
signals with X-ray emission lagging optical one by a few seconds 
and narrower optical auto-correlation functions 
(ACFs) to X-ray ones in several LMXBs 
\citep[e.g.,][]{gan08gx339,dur08swiftj1753,mal04xtej1118model,vel11sscmodel}.  
The timescales of the X-ray delays detected in this study, 
however, are inconsistent with the odd timing properties 
expected by these models as well as the smoother optical flares 
to the X-ray ones and the positive peaks at negative optical 
time lags in the DCFs (see also Figures \ref{datacorr} and S2).  

   There is some evidence to support that the 
effect of X-ray reprocessing and/or cyclo-synchrotron emission 
was weak.  
First, the estimated relation between the optical and X-ray 
luminosity in Sec.~3.2 was $L_{\rm opt} \propto 
L_{\rm X}^{0.25-0.29}$ and this value disagrees with both 
that expected by standard disc reprocessing \citep[$L_{\rm V} 
\propto L_{\rm X}^{0.5}$,][]{van94visualLMXB} and that predicted 
by jet ejections plus X-ray irradiation \citep[$L_{\rm opt} 
\propto L_{\rm X}^{0.5-0.7}$,][]{rus06OIRandXrayCorr}.  
Second, the optical/X-ray flux ratios in the December outburst 
($\sim$0.05 in the $V$ band and $\sim$0.01 in the $I_{\rm C}$ 
band) were smaller by a factor of 2--5 than those in the June 
outburst 
when X-ray reprocessing considerably dominated the optical flux.  
This would indicate that the effect of reprocessing was weaker 
in the December outburst than that in the June outburst.  
We suggest the reason is that the outer disc would 
be depleted due to ionisation during the June outburst and/or 
the strong outflow discussed in \citet{mun16v404cyg}.  On the 
other hand, in the June outburst, the outer disc is thought 
to have been optically thick \citep{kim16v404cyg}.

\subsection{Time Delay Caused by Propagation of Mass Accretion Flow in the Inner Disc}

   We propose an interpretation that short-term variations of 
the mass-accretion rate in the outer disc (whose origin is still 
unknown) propagate to the inner disc via optical fluctuations 
which will thereby prompt X-ray fluctuations.  This then 
replaces the mechanisms discussed in the previous subsection 
as a possible origin of the delays.  
   If the standard thin disc extends close to the central black 
hole, it will take longer for the observed delays to propagate 
in an accretion flow from the optical emission region to the X-ray 
emission region.  This is because the speed of a propagating heating 
wave is proportional to the speed of sound ($c_{\rm s}$) in the 
standard disc \citep[i.e., $v_{\rm f} \sim \alpha 
c_{\rm s}$;][]{mey84ADtransitionwave}.
Here, $\alpha$ represents the viscous parameter.  
Thus we consider the condition that the disc is composed of 
an optically-thin flow as an advection-dominated accretion 
flow (ADAF) and a truncated geometrically-thin standard disc.  
This picture was considered during the 2002 outburst in V4641 
Sgr, another black-hole transient LMXB, when a 7-min delay of 
the X-ray variations against the optical ones was detected at 
the fading stage \citep{uem04v4641sgr}.  

   We assume that the thin standard disc extended to the 
transition radius ($R_{\rm tr}$) and that there was the ADAF 
on the inside of the radius \citep[e.g.,][]{ham97j1655ADAF}.  
In the ADAF, the matter moves to the central object with 
a speed comparable to the free-fall velocity ($v_{\rm ff}$) 
\citep[$v_{\rm f} \sim \alpha v_{\rm ff}$;][]{nar95ADAF}.  
If the optical fluctuations, which were 
triggered at a region close to the truncation radius, 
propagated to the central black hole via the accretion of 
mass flow on the free-fall timescale in the ADAF, the transition 
radius is estimated to be $\sim$2.5--4.0$\times 10^{8}$ [cm] 
by using the above approximation for $\alpha = 0.01$ 
\citep{san98localMHD,mac02globalMHD} and our estimates of 
X-ray delays.  
The estimated value of $R_{\rm tr}$ corresponds to 
$\sim$100--150$r_{\rm s}$.  
Here, $r_{\rm s}$ ($= 2GM/c^{2}$) represents the Schwarzschild 
radius for a 9$M_{\odot}$ black hole.  
The estimated value is close to that of the inner disc radius 
derived through the SED analyses in the June outburst \citep[see 
Sec.~8 of Methods in][]{kim16v404cyg}.  The region at the radius 
is, however, too hot to emit thermal optical photons predominantly; 
hence the picture may be problematic.  It is likely that the 
optical emission was non-thermal as discussed in e.g., 
\citet{uem02v4641sgrletter}.  
   In addition, the smaller amplitude of the optical variations 
can be explained by the presence of the optical continuum emission 
from the outer disc.

\subsection{Testing Possibilities of the Origin of X-ray Delays}

   An optically-thin flow like an ADAF inside a truncated 
standard disc has been widely proposed as 
a possible scenario for the low/hard state in 
black-hole X-ray binaries, although the interpretation 
remains under discussion 
\citep[][for a review]{rem06BHBreview,don07XB,bel11BHtransients}.  
   The average value of the bolometric luminosity during 
intervals (1) and (2), 0.03--0.1$L_{\rm Edd}$, 
is consistent with the low/hard state 
\citep{don03eventhorizon,dun08BHspectra}, 
the existing theory \citep{yua07xtej1550model} and the 3D 
magnetohydrodynamics (MHD) simulations 
\citep{mat06HStransition,oda07magretoBHdisk}.  
   The spectral state was also reported to be the low/hard 
state during the day 8.15--10.32 \citep{mot16atel8500}; 
however, a detailed spectral analysis would be useful to 
test our interpretation since the peculiar X-ray spectral 
behaviour similar to that of an obscured AGN was found 
for some time intervals during the June 
outburst by \citet{san17v404cyg} and \citet{mot17v404cyg}.  
They suggested, on the basis of the spectral 
behaviour, that optically thick outflowing material would 
obscure substantial X-rays from the central part of V404 
Cyg and that the intrinsic luminosity was close to the Eddington 
luminosity.  
This situation implies the existence of a slim disc; however, 
it would be difficult for short-term variability and time 
delays to coexist in this circumstance.\footnote{Sporadic 
mass accretion is unlikely to occur due to the high accretion 
rate in a slim disc.  
A variable absorber to cause the observed variability in 
the June outburst was suggested by 
\citet{san17v404cyg} and \citet{mot17v404cyg} instead, but 
this condition would not produce time lags between optical 
and X-ray emission.  }
Additionally, a slim disc itself would be difficult to 
maintain during the December outburst since the averaged 
X-ray flux during this outburst was much smaller than that 
during the June outburst.  

\section{Conclusions}

   We report on the photometric observations in the 
outburst of V404 Cyg from December, 2015 to January, 2016.  
Our main findings are summarised below.  

\begin{enumerate}
\item
The 2015 December outburst in V404 Cyg was very similar to 
the 2015 June--July outburst in the following two aspects.  
One is that violent and rapid optical variability with 
amplitudes of $\sim$2 mag on timescales of 
$\sim$10 min--3 hours was observed at low luminosity.  
It is likely that this kind of variation is commonly seen 
in the outbursts including the 1989 one in this object.  
The other is that the trend of the overall light curves was 
a slow rise and rapid decay.  
\item
We detected an X-ray delay of $\sim$30--50 s 
against the optical emission in the two intervals during the 
December outburst, when the large-amplitude stochastic optical 
variations were observed at the low average luminosity, 
$\sim$0.03--0.1$L_{\rm Edd}$.  
In addition, the relation between the optical and X-ray luminosity 
for these two intervals was $L_{\rm opt} \propto L_{\rm X}^{0.25-0.29}$.  
We suggest that the X-ray delay can be due to propagation of 
mass accretion flow in an inner optically-thin hot flow like 
an ADAF with a speed comparable to the free-fall velocity.  
\end{enumerate}

\section*{Acknowledgements}

We acknowledge the variable star observations from the AAVSO 
International Database contributed by observers worldwide and 
used in this research.  We also thank the {\it INTEGRAL} groups 
for making the products of the ToO data publicly available online 
at the {\it INTEGRAL} Science Data Centre.  This work was 
financially supported by the Grant-in-Aid for JSPS Fellows 
for young researchers (MK) and 
the Grant-in-Aid ``Initiative for High-Dimensional Data-Driven 
Science through Deepening of Sparse Modeling'' from the Ministry 
of Education, Culture, Sports, Science and Technology (MEXT) of 
Japan (25120007, TK).  
It was also partially supported by the RFBR grant 15-02-06178.  
We are thankful to many amateur observers for providing a lot of 
data used in this research.  
Hyungsuk Tak acknowledges partial support from the United States 
National Science Foundation under Grant DMS 1127914 to the SAMSI.  
We give special thanks to Kaisey Mandel and David van Dyk for 
very helpful discussions.  
Yutaro Tachibana provided us with a code for calculations of LNDCFs.  
Roger D.~Pickard gratefully acknowledges use of the Las Cumbres 
Observatory Global Telescope Network.  
We are grateful to an anonymous referee for his/her helpful comments.  






\begin{thebibliography}{}
\makeatletter
\relax
\def\mn@urlcharsother{\let\do\@makeother \do\$\do\&\do\#\do\^\do\_\do\%\do\~}
\def\mn@doi{\begingroup\mn@urlcharsother \@ifnextchar [ {\mn@doi@}
  {\mn@doi@[]}}
\def\mn@doi@[#1]#2{\def\@tempa{#1}\ifx\@tempa\@empty \href
  {http://dx.doi.org/#2} {doi:#2}\else \href {http://dx.doi.org/#2} {#1}\fi
  \endgroup}
\def\mn@eprint#1#2{\mn@eprint@#1:#2::\@nil}
\def\mn@eprint@arXiv#1{\href {http://arxiv.org/abs/#1} {{\tt arXiv:#1}}}
\def\mn@eprint@dblp#1{\href {http://dblp.uni-trier.de/rec/bibtex/#1.xml}
  {dblp:#1}}
\def\mn@eprint@#1:#2:#3:#4\@nil{\def\@tempa {#1}\def\@tempb {#2}\def\@tempc
  {#3}\ifx \@tempc \@empty \let \@tempc \@tempb \let \@tempb \@tempa \fi \ifx
  \@tempb \@empty \def\@tempb {arXiv}\fi \@ifundefined
  {mn@eprint@\@tempb}{\@tempb:\@tempc}{\expandafter \expandafter \csname
  mn@eprint@\@tempb\endcsname \expandafter{\@tempc}}}

\bibitem[\protect\citeauthoryear{{Barthelmy}, {D'Ai}, {D'Avanzo}, {Krimm},
  {Lien}, {Marshall}, {Maselli}  \& {Siegel}}{{Barthelmy}
  et~al.}{2015a}]{bar15v404cyggcn17929}
{Barthelmy} S.~D.,  {D'Ai} A.,  {D'Avanzo} P.,  {Krimm} H.~A.,  {Lien} A.~Y.,
  {Marshall} F.~E.,  {Maselli} A.,   {Siegel} M.~H.,  2015a, GRB Coordinates
  Network, 17929

\bibitem[\protect\citeauthoryear{{Barthelmy}, {Page}  \& {Palmer}}{{Barthelmy}
  et~al.}{2015b}]{bar15bGCNv404cyg}
{Barthelmy} S.~D.,  {Page} K.~L.,   {Palmer} D.~M.,  2015b, GRB Coordinates
  Network, 18716

\bibitem[\protect\citeauthoryear{{Belloni}, {Motta}  \&
  {Mu{\~n}oz-Darias}}{{Belloni} et~al.}{2011}]{bel11BHtransients}
{Belloni} T.~M.,  {Motta} S.~E.,   {Mu{\~n}oz-Darias} T.,  2011, Bulletin of
  the Astronomical Society of India, 39, 409

\bibitem[\protect\citeauthoryear{{Bernardini}, {Russell}, {Shaw}, {Lewis},
  {Charles}, {Koljonen}, {Lasota}  \& {Casares}}{{Bernardini}
  et~al.}{2016a}]{ber16v404cyg}
{Bernardini} F.,  {Russell} D.~M.,  {Shaw} A.~W.,  {Lewis} F.,  {Charles}
  P.~A.,  {Koljonen} K.~I.~I.,  {Lasota} J.~P.,   {Casares} J.,  2016a, \apjl, 
818, L5

\bibitem[\protect\citeauthoryear{{Bernardini}, {Russell}, {Kolojonen},
  {Stella}, {Hynes}  \& {Corbel}}{{Bernardini}
  et~al.}{2016b}]{ber16cenX4v404cyg}
{Bernardini} F.,  {Russell} D.~M.,  {Kolojonen} K.~I.~I.,  {Stella} L.,
  {Hynes} R.~I.,   {Corbel} S.,  2016b, \apj, 
  826, 149

\bibitem[\protect\citeauthoryear{{Cardelli}, {Clayton}  \& {Mathis}}{{Cardelli}
  et~al.}{1989}]{car89extinction}
{Cardelli} J.~A.,  {Clayton} G.~C.,   {Mathis} J.~S.,  1989, \apj, 
  345, 245

\bibitem[\protect\citeauthoryear{{Casares}, {Charles}, {Naylor}  \&
  {Pavlenko}}{{Casares} et~al.}{1993}]{cas93v404cyg}
{Casares} J.,  {Charles} P.~A.,  {Naylor} T.,   {Pavlenko} E.~P.,  1993,
  \mnras, 265, 834

\bibitem[\protect\citeauthoryear{Chen, Shrader  \& Livio}{Chen
  et~al.}{1997}]{che97BHXN}
Chen W.,  Shrader C.~R.,   Livio M.,  1997, \apj, 491, 312

\bibitem[\protect\citeauthoryear{{Done} \& {Gierli{\'n}ski}}{{Done} \&
  {Gierli{\'n}ski}}{2003}]{don03eventhorizon}
{Done} C.,  {Gierli{\'n}ski} M.,  2003, \mnras, 
342, 1041

\bibitem[\protect\citeauthoryear{{Done}, {Gierli{\'n}ski}  \& {Kubota}}{{Done}
  et~al.}{2007}]{don07XB}
{Done} C.,  {Gierli{\'n}ski} M.,   {Kubota} A.,  2007, \aapr, 15, 1

\bibitem[\protect\citeauthoryear{Dubus, Hameury  \& Lasota}{Dubus
  et~al.}{2001}]{dub01XNmodel}
Dubus G.,  Hameury J.-M.,   Lasota J.-P.,  2001, \aap, 373, 251

\bibitem[\protect\citeauthoryear{{Dunn}, {Fender}, {K{\"o}rding}, {Belloni}  \&
  {Cabanac}}{{Dunn} et~al.}{2010}]{dun08BHspectra}
{Dunn} R.~J.~H.,  {Fender} R.~P.,  {K{\"o}rding} E.~G.,  {Belloni} T.,
  {Cabanac} C.,  2010, \mnras, 
  403, 61

\bibitem[\protect\citeauthoryear{{Durant}, {Gandhi}, {Shahbaz}, {Fabian},
  {Miller}, {Dhillon}  \& {Marsh}}{{Durant} et~al.}{2008}]{dur08swiftj1753}
{Durant} M.,  {Gandhi} P.,  {Shahbaz} T.,  {Fabian} A.~P.,  {Miller} J.,
  {Dhillon} V.~S.,   {Marsh} T.~R.,  2008, \apjl, 682, L45

\bibitem[\protect\citeauthoryear{{Gandhi} et~al.,}{{Gandhi}
  et~al.}{2008}]{gan08gx339}
{Gandhi} P.,  et~al., 2008, \mnras, 390, L29

\bibitem[\protect\citeauthoryear{{Gandhi} et~al.,}{{Gandhi}
  et~al.}{2016}]{gan16v404cyg}
{Gandhi} P.,  et~al., 2016, \mnras, 459, 554

\bibitem[\protect\citeauthoryear{{Gelman} \& {Rubin}}{{Gelman} \&
  {Rubin}}{1992}]{gelmanrubin92}
{Gelman} A.,  {Rubin} D.~B.,  1992, Statistical Science., pp 457--472

\bibitem[\protect\citeauthoryear{Hameury, Lasota, McClintock  \&
  Narayan}{Hameury et~al.}{1997}]{ham97j1655ADAF}
Hameury J.-M.,  Lasota J.-P.,  McClintock J.~E.,   Narayan R.,  1997, \apj,
  489, 234

\bibitem[\protect\citeauthoryear{{Huppenkothen} et~al.,}{{Huppenkothen}
  et~al.}{2017}]{hup17v404cyg}
{Huppenkothen} D.,  et~al., 2017, \apj, 834, 90

\bibitem[\protect\citeauthoryear{Hynes, O'Brien, Horne, Chen  \& Haswell}{Hynes
  et~al.}{1998}]{hyn98j1655echo}
Hynes R.~I.,  O'Brien K.,  Horne K.,  Chen W.,   Haswell C.~A.,  1998, \mnras,
  299, 37P

\bibitem[\protect\citeauthoryear{{Hynes} et~al.,}{{Hynes}
  et~al.}{2004}]{hyn04v404cyg}
{Hynes} R.~I.,  et~al., 2004, \apjl, 611, L125

\bibitem[\protect\citeauthoryear{{Hynes}, {Bradley}, {Rupen}, {Gallo},
  {Fender}, {Casares}  \& {Zurita}}{{Hynes} et~al.}{2009}]{hyn09v404cyg}
{Hynes} R.~I.,  {Bradley} C.~K.,  {Rupen} M.,  {Gallo} E.,  {Fender} R.~P.,
  {Casares} J.,   {Zurita} C.,  2009, \mnras, 399, 2239

\bibitem[\protect\citeauthoryear{{Jenke} et~al.,}{{Jenke}
  et~al.}{2016}]{jen16v404cyg}
{Jenke} P.~A.,  et~al., 2016, \apj, 826, 37

\bibitem[\protect\citeauthoryear{{Jourdain}, {Roques}  \& {Rodi}}{{Jourdain}
  et~al.}{2017}]{jou17v404cyg}
{Jourdain} E.,  {Roques} J.-P.,   {Rodi} J.,  2017, \apj, 
  834, 130

\bibitem[\protect\citeauthoryear{{Kanbach}, {Straubmeier}, {Spruit}  \&
  {Belloni}}{{Kanbach} et~al.}{2001}]{kan01j1118varcorrelation}
{Kanbach} G.,  {Straubmeier} C.,  {Spruit} H.~C.,   {Belloni} T.,  2001, \nat,
  414, 180

\bibitem[\protect\citeauthoryear{Kato, Uemura, Ishioka, Nogami, Kunjaya, Baba
  \& Yamaoka}{Kato et~al.}{2004}]{VSNET}
Kato T.,  Uemura M.,  Ishioka R.,  Nogami D.,  Kunjaya C.,  Baba H.,   Yamaoka
  H.,  2004, \pasj, 56, S1

\bibitem[\protect\citeauthoryear{{Kato}, {Fukue}  \& {Mineshige}}{{Kato}
  et~al.}{2008}]{kat08BHaccretion}
{Kato} S.,  {Fukue} J.,   {Mineshige} S.,  2008, Kyoto University Press (Kyoto, Japan), {Black-Hole Accretion Disks
  --- Towards a New Paradigm ---}

\bibitem[\protect\citeauthoryear{{Kelly}, {Bechtold}  \&
  {Siemiginowska}}{{Kelly} et~al.}{2009}]{kel09QuasarOptical}
{Kelly} B.~C.,  {Bechtold} J.,   {Siemiginowska} A.,  2009, \apj, 
  698, 895

\bibitem[\protect\citeauthoryear{{Khargharia}, {Froning}  \&
  {Robinson}}{{Khargharia} et~al.}{2010}]{kha10v404cygcenx4}
{Khargharia} J.,  {Froning} C.~S.,   {Robinson} E.~L.,  2010, \apj, 716, 1105

\bibitem[\protect\citeauthoryear{{Kimura} et~al.,}{{Kimura}
  et~al.}{2016}]{kim16v404cyg}
{Kimura} M.,  et~al., 2016, \nat, 529, 54

\bibitem[\protect\citeauthoryear{{King}, {Miller}, {Raymond}, {Reynolds}  \&
  {Morningstar}}{{King} et~al.}{2015}]{kin15v404cyg}
{King} A.~L.,  {Miller} J.~M.,  {Raymond} J.,  {Reynolds} M.~T.,
  {Morningstar} W.,  2015, \apjl, 813, L37

\bibitem[\protect\citeauthoryear{{Koz{\l}owski} et~al.,}{{Koz{\l}owski}
  et~al.}{2010}]{koz10QuasarVari}
{Koz{\l}owski} S.,  et~al., 2010, \apj, 708, 927

\bibitem[\protect\citeauthoryear{{Kuulkers}, {amp}  \& {Ferrigno}}{{Kuulkers}
  et~al.}{2016}]{kul16atel8512}
{Kuulkers} E.,  {amp}  {Ferrigno} C.,  2016, The Astronomer's Telegram, 8512

\bibitem[\protect\citeauthoryear{{Lasota}}{{Lasota}}{2001}]{las01DIDNXT}
{Lasota} J.-P.,  2001, \nar, 45, 449

\bibitem[\protect\citeauthoryear{{Lehar}, {Hewitt}, {Burke}  \&
  {Roberts}}{{Lehar} et~al.}{1992}]{leh92timedelay}
{Lehar} J.,  {Hewitt} J.~N.,  {Burke} B.~F.,   {Roberts} D.~H.,  1992, \apj, 384, 453

\bibitem[\protect\citeauthoryear{{Lipunov} et~al.,}{{Lipunov}
  et~al.}{2015}]{lip15bv404cyg}
{Lipunov} V.,  et~al., 2015, The Astronomer's Telegram, 8453

\bibitem[\protect\citeauthoryear{{Lipunov} et~al.,}{{Lipunov}
  et~al.}{2016}]{lip16v404cyg}
{Lipunov} V.~M.,  et~al., 2016, \apj, 833, 198

\bibitem[\protect\citeauthoryear{{Loh} et~al.,}{{Loh}
  et~al.}{2016}]{loh16v404cyg}
{Loh} A.,  et~al., 2016, \mnras, 462, L111

\bibitem[\protect\citeauthoryear{{MacLeod} et~al.,}{{MacLeod}
  et~al.}{2010}]{mac10DRW}
{MacLeod} C.~L.,  et~al., 2010, \apj, 721, 1014

\bibitem[\protect\citeauthoryear{{Machida} \& {Matsumoto}}{{Machida} \&
  {Matsumoto}}{2003}]{mac02globalMHD}
{Machida} M.,  {Matsumoto} R.,  2003, \apj, 585, 429

\bibitem[\protect\citeauthoryear{{Machida}, {Nakamura}  \&
  {Matsumoto}}{{Machida} et~al.}{2006}]{mat06HStransition}
{Machida} M.,  {Nakamura} K.~E.,   {Matsumoto} R.,  2006, \pasj, 
  {10.1093/pasj/58.1.193}, 58, 193

\bibitem[\protect\citeauthoryear{{Makino}}{{Makino}}{1989}]{mak89v404cygiauc4782}
{Makino} F.,  1989, \iaucirc, 4782

\bibitem[\protect\citeauthoryear{{Malzac}, {Merloni}  \& {Fabian}}{{Malzac}
  et~al.}{2004}]{mal04xtej1118model}
{Malzac} J.,  {Merloni} A.,   {Fabian} A.~C.,  2004, \mnras, 
351, 253

\bibitem[\protect\citeauthoryear{Markoff, Falcke  \& Fender}{Markoff
  et~al.}{2001}]{mar01j1118}
Markoff S.,  Falcke H.,   Fender R.,  2001, \aap, 372, L25

\bibitem[\protect\citeauthoryear{{Mart{\'{\i}}}, {Luque-Escamilla}  \&
  {Garc{\'{\i}}a-Hern{\'a}ndez}}{{Mart{\'{\i}}} et~al.}{2016}]{mar16v404cyg}
{Mart{\'{\i}}} J.,  {Luque-Escamilla} P.~L.,   {Garc{\'{\i}}a-Hern{\'a}ndez}
  M.~T.,  2016, \aap, 586, A58

\bibitem[\protect\citeauthoryear{Merloni, Di~Matteo  \& Fabian}{Merloni
  et~al.}{2000}]{mer00j1118}
Merloni A.,  Di~Matteo T.,   Fabian A.~C.,  2000, \mnras, 318, L15

\bibitem[\protect\citeauthoryear{{Meyer}}{{Meyer}}{1984}]{mey84ADtransitionwave}
{Meyer} F.,  1984, \aap, 131, 303

\bibitem[\protect\citeauthoryear{{Miller-Jones}, {Jonker}, {Dhawan}, {Brisken},
  {Rupen}, {Nelemans}  \& {Gallo}}{{Miller-Jones}
  et~al.}{2009}]{mil09v404cygdistance}
{Miller-Jones} J.~C.~A.,  {Jonker} P.~G.,  {Dhawan} V.,  {Brisken} W.,  {Rupen}
  M.~P.,  {Nelemans} G.,   {Gallo} E.,  2009, \apjl, 706, L230

\bibitem[\protect\citeauthoryear{{Mirabel}, {Dhawan}, {Chaty}, {Rodriguez},
  {Marti}, {Robinson}, {Swank}  \& {Geballe}}{{Mirabel}
  et~al.}{1998}]{mir98grs1915}
{Mirabel} I.~F.,  {Dhawan} V.,  {Chaty} S.,  {Rodriguez} L.~F.,  {Marti} J.,
  {Robinson} C.~R.,  {Swank} J.,   {Geballe} T.,  1998, \aap, 330, L9

\bibitem[\protect\citeauthoryear{{Mooley}, {Fender}, {Anderson}, {Staley},
  {Kuulkers}  \& {Rumsey}}{{Mooley} et~al.}{2015}]{moo15atel7658}
{Mooley} K.,  {Fender} R.,  {Anderson} G.,  {Staley} T.,  {Kuulkers} E.,
  {Rumsey} C.,  2015, The Astronomer's Telegram, 7658

\bibitem[\protect\citeauthoryear{Motch, Ilovaisky  \& Chevalier}{Motch
  et~al.}{1982}]{mot82gx339}
Motch C.,  Ilovaisky S.~A.,   Chevalier C.,  1982, \aap, 109, L1

\bibitem[\protect\citeauthoryear{{Motta}, {Sanchez-Fernandez}, {Kuulkers},
  {Kajava}  \& {Bozzo}}{{Motta} et~al.}{2016}]{mot16atel8500}
{Motta} S.~E.,  {Sanchez-Fernandez} C.,  {Kuulkers} E.,  {Kajava} J.,   {Bozzo}
  E.,  2016, The Astronomer's Telegram, 8500

\bibitem[\protect\citeauthoryear{{Motta}, {Kajava},
  {S{\'a}nchez-Fern{\'a}ndez}, {Giustini}  \& {Kuulkers}}{{Motta}
  et~al.}{2017}]{mot17v404cyg}
{Motta} S.~E.,  {Kajava} J.~J.~E.,  {S{\'a}nchez-Fern{\'a}ndez} C.,  {Giustini}
  M.,   {Kuulkers} E.,  2017, \mnras, 468, 981

\bibitem[\protect\citeauthoryear{{Mu{\~n}oz-Darias} et~al.,}{{Mu{\~n}oz-Darias}
  et~al.}{2016}]{mun16v404cyg}
{Mu{\~n}oz-Darias} T.,  et~al., 2016, \nat, 534, 75

\bibitem[\protect\citeauthoryear{{Mu{\~n}oz-Darias} et~al.,}{{Mu{\~n}oz-Darias}
  et~al.}{2017}]{mun16v404cyg2}
{Mu{\~n}oz-Darias} T.,  et~al., 2017, \mnras, 465, L124

\bibitem[\protect\citeauthoryear{{Narayan} \& {Yi}}{{Narayan} \&
  {Yi}}{1995}]{nar95ADAF}
{Narayan} R.,  {Yi} I.,  1995, \apj, 452, 710

\bibitem[\protect\citeauthoryear{{Natalucci}, {Fiocchi}, {Bazzano}, {Ubertini},
  {Roques}  \& {Jourdain}}{{Natalucci} et~al.}{2015}]{nat15v404cyg}
{Natalucci} L.,  {Fiocchi} M.,  {Bazzano} A.,  {Ubertini} P.,  {Roques} J.-P.,
   {Jourdain} E.,  2015, \apjl, 813, L21

\bibitem[\protect\citeauthoryear{{Negoro} et~al.,}{{Negoro}
  et~al.}{2015}]{neg15atel7646}
{Negoro} H.,  et~al., 2015, The Astronomer's Telegram, 7646

\bibitem[\protect\citeauthoryear{{Oda}, {Machida}, {Nakamura}  \&
  {Matsumoto}}{{Oda} et~al.}{2007}]{oda07magretoBHdisk}
{Oda} H.,  {Machida} M.,  {Nakamura} K.~E.,   {Matsumoto} R.,  2007, \pasj, 59, 457

\bibitem[\protect\citeauthoryear{{Pelt}, {Hoff}, {Kayser}, {Refsdal}  \&
  {Schramm}}{{Pelt} et~al.}{1994}]{pel94timedelay}
{Pelt} J.,  {Hoff} W.,  {Kayser} R.,  {Refsdal} S.,   {Schramm} T.,  1994,
  \aap, 286, 775

\bibitem[\protect\citeauthoryear{{Piano}, {Munar-Adrover}, {Verrecchia},
  {Tavani}  \& {Trushkin}}{{Piano} et~al.}{2017}]{pia17v404cyg}
{Piano} G.,  {Munar-Adrover} P.,  {Verrecchia} F.,  {Tavani} M.,   {Trushkin}
  S.~A.,  2017, \apj, 839, 84

\bibitem[\protect\citeauthoryear{{Radhika}, {Nandi}, {Agrawal}  \&
  {Mandal}}{{Radhika} et~al.}{2016}]{rad16v404cyg}
{Radhika} D.,  {Nandi} A.,  {Agrawal} V.~K.,   {Mandal} S.,  2016, \mnras, 
462, 1834

\bibitem[\protect\citeauthoryear{{Remillard} \& {McClintock}}{{Remillard} \&
  {McClintock}}{2006}]{rem06BHBreview}
{Remillard} R.~A.,  {McClintock} J.~E.,  2006, \araa, 44, 49

\bibitem[\protect\citeauthoryear{{Rodriguez} et~al.,}{{Rodriguez}
  et~al.}{2015}]{rod15v404cyg}
{Rodriguez} J.,  et~al., 2015, \aap, 581, L9

\bibitem[\protect\citeauthoryear{{Roques} \& {Jourdain}}{{Roques} \&
  {Jourdain}}{2016}]{roq16v404cyg}
{Roques} J.~P.,  {Jourdain} E.,  2016, preprint, (arXiv:1601.05289)

\bibitem[\protect\citeauthoryear{{Russell}, {Fender}, {Hynes}, {Brocksopp},
  {Homan}, {Jonker}  \& {Buxton}}{{Russell} et~al.}{2006}]{rus06OIRandXrayCorr}
{Russell} D.~M.,  {Fender} R.~P.,  {Hynes} R.~I.,  {Brocksopp} C.,  {Homan} J.,
   {Jonker} P.~G.,   {Buxton} M.~M.,  2006, \mnras, 371, 1334

\bibitem[\protect\citeauthoryear{{Sanchez-Fernandez}, {Kajava}, {Motta}  \&
  {Kuulkers}}{{Sanchez-Fernandez} et~al.}{2017}]{san17v404cyg}
{Sanchez-Fernandez} C.,  {Kajava} J.~J.~E.,  {Motta} S.~E.,   {Kuulkers} E.,
  2017, \aap, 602, 40

\bibitem[\protect\citeauthoryear{{Sano}, {Inutsuka}  \& {Miyama}}{{Sano}
  et~al.}{1998}]{san98localMHD}
{Sano} T.,  {Inutsuka} S.-i.,   {Miyama} S.~M.,  1998, \apjl, 
506, L57

\bibitem[\protect\citeauthoryear{{Segreto}, {Del Santo}, {D'A{\'{\i}}}, {La
  Parola}, {Cusumano}, {Mineo}  \& {Malzac}}{{Segreto}
  et~al.}{2015}]{seg15atel7755}
{Segreto} A.,  {Del Santo} M.,  {D'A{\'{\i}}} A.,  {La Parola} V.,  {Cusumano}
  G.,  {Mineo} T.,   {Malzac} J.,  2015, The Astronomer's Telegram, 7755

\bibitem[\protect\citeauthoryear{{Shahbaz}, {Ringwald}, {Bunn}, {Naylor},
  {Charles}  \& {Casares}}{{Shahbaz} et~al.}{1994}]{sha93v404cyg}
{Shahbaz} T.,  {Ringwald} F.~A.,  {Bunn} J.~C.,  {Naylor} T.,  {Charles} P.~A.,
    {Casares} J.,  1994, \mnras, 271, L10

\bibitem[\protect\citeauthoryear{{Shahbaz}, {Russell}, {Covino}, {Mooley},
  {Fender}  \& {Rumsey}}{{Shahbaz} et~al.}{2016}]{sha16v404cyg}
{Shahbaz} T.,  {Russell} D.~M.,  {Covino} S.,  {Mooley} K.,  {Fender} R.~P.,
  {Rumsey} C.,  2016, \mnras, 463, 1822

\bibitem[\protect\citeauthoryear{{Shakura} \& {Sunyaev}}{{Shakura} \&
  {Sunyaev}}{1973}]{sha73BHbinary}
{Shakura} N.~I.,  {Sunyaev} R.~A.,  1973, \aap, 24, 337

\bibitem[\protect\citeauthoryear{{Siegert} et~al.,}{{Siegert}
  et~al.}{2016}]{sie16v404cyg}
{Siegert} T.,  et~al., 2016, \nat, 531, 341

\bibitem[\protect\citeauthoryear{{Tak}, {Mandel}, {van Dyk}, {Kashyap}, {Meng}
  \& {Siemiginowska}}{{Tak} et~al.}{2016a}]{tak16timedelay}
{Tak} H.,  {Mandel} K.,  {van Dyk} D.~A.,  {Kashyap} V.~L.,  {Meng} X.-L.,
  {Siemiginowska} A.,  2016a, preprint, (arXiv:1602.01462)

\bibitem[\protect\citeauthoryear{Tak, Meng  \& van Dyk}{Tak
  et~al.}{2016b}]{tak16multi}
Tak H.,  Meng X.-L.,   van Dyk D.~A.,  2016b, preprint, (arXiv:1601.05633)

\bibitem[\protect\citeauthoryear{{Tanaka} \& {Shibazaki}}{{Tanaka} \&
  {Shibazaki}}{1996}]{tan96XNreview}
{Tanaka} Y.,  {Shibazaki} N.,  1996, \araa, 34, 607

\bibitem[\protect\citeauthoryear{{Tanaka} et~al.,}{{Tanaka}
  et~al.}{2016}]{tan16v404cyg}
{Tanaka} Y.~T.,  et~al., 2016, \apj, 823, 35

\bibitem[\protect\citeauthoryear{{Tetarenko}, {Sivakoff}, {Young}, {Wouterloot}
   \& {Miller-Jones}}{{Tetarenko} et~al.}{2015}]{tet15v404cygatel7708}
{Tetarenko} A.,  {Sivakoff} G.~R.,  {Young} K.,  {Wouterloot} J.~G.~A.,
  {Miller-Jones} J.~C.,  2015, The Astronomer's Telegram, 7708

\bibitem[\protect\citeauthoryear{Tierney}{Tierney}{1994}]{tie94markov}
Tierney L.,  1994, The Annals of Statistics, 22, 1701

\bibitem[\protect\citeauthoryear{Uemura et~al.,}{Uemura
  et~al.}{2002}]{uem02v4641sgrletter}
Uemura M.,  et~al., 2002, \pasj, 54, L79

\bibitem[\protect\citeauthoryear{{Uemura} et~al.,}{{Uemura}
  et~al.}{2004}]{uem04v4641sgr}
{Uemura} M.,  et~al., 2004, \pasj, 56, S61

\bibitem[\protect\citeauthoryear{{Veledina}, {Poutanen}  \& {Vurm}}{{Veledina}
  et~al.}{2011}]{vel11sscmodel}
{Veledina} A.,  {Poutanen} J.,   {Vurm} I.,  2011, \apjl, 737, L17

\bibitem[\protect\citeauthoryear{Wagner, Starrfield, Cassatella, Hurst,
  Mobberley  \& Marsden}{Wagner et~al.}{1989}]{wag89v404cygiauc4783}
Wagner R.~M.,  Starrfield S.~G.,  Cassatella A.,  Hurst G.~M.,  Mobberley M.,
  Marsden B.~G.,  1989, \iaucirc, 4783

\bibitem[\protect\citeauthoryear{{Wagner}, {Bertram}, {Starrfield}, {Howell},
  {Kreidl}, {Bus}, {Cassatella}  \& {Fried}}{{Wagner}
  et~al.}{1991}]{wag91v404cyg}
{Wagner} R.~M.,  {Bertram} R.,  {Starrfield} S.~G.,  {Howell} S.~B.,  {Kreidl}
  T.~J.,  {Bus} S.~J.,  {Cassatella} A.,   {Fried} R.,  1991, \apj, 378, 293

\bibitem[\protect\citeauthoryear{Wagner, Kreidl, Howell  \& Starrfield}{Wagner
  et~al.}{1992}]{wag92v404cyg}
Wagner R.~M.,  Kreidl T.~J.,  Howell S.~B.,   Starrfield S.~G.,  1992, \apjl,
  401, L97

\bibitem[\protect\citeauthoryear{{Walton} et~al.,}{{Walton}
  et~al.}{2017}]{wal17v404cyg}
{Walton} D.~J.,  et~al., 2017, \apj, 839, 110

\bibitem[\protect\citeauthoryear{{Yuan}, {Zdziarski}, {Xue}  \& {Wu}}{{Yuan}
  et~al.}{2007}]{yua07xtej1550model}
{Yuan} F.,  {Zdziarski} A.~A.,  {Xue} Y.,   {Wu} X.-B.,  2007, \apj, 
659, 541

\bibitem[\protect\citeauthoryear{{{\.Z}ycki}, {Done}  \& {Smith}}{{{\.Z}ycki}
  et~al.}{1999}]{zyc99v404cyg}
{{\.Z}ycki} P.~T.,  {Done} C.,   {Smith} D.~A.,  1999, \mnras, 309, 561

\bibitem[\protect\citeauthoryear{{van Paradijs} \& {McClintock}}{{van Paradijs}
  \& {McClintock}}{1994}]{van94visualLMXB}
{van Paradijs} J.,  {McClintock} J.~E.,  1994, \aap, 290, 133

\bibitem[\protect\citeauthoryear{{van der Laan}}{{van der
  Laan}}{1966}]{van66RadioModel}
{van der Laan} H.,  1966, \nat, 211, 1131

\makeatother
\end{thebibliography}

\newcommand{\noop}[1]{}

\renewcommand{\thetable}{%
  S\arabic{table}}
\renewcommand{\thefigure}{%
  S\arabic{figure}}

\setcounter{table}{0}
\setcounter{figure}{0}

\begin{table*}
	\centering
	\caption{List of Instruments used for the photometry of the 2015 December outburst in V404 Cyg.}
	\label{telescope}
	\begin{tabular}{lccc}
		\hline
		CODE$^{*}$ & Telescope (\& CCD) & Observatory (or Observer) & Site\\
		\hline
COO & T07$^{\dagger}$ 43cm+STL-1100M & AstroCamp Observatory & Nerpio, Spain \\
    & T21$^{\dagger}$ 43cm+FLI-PL6303E & iTelescope.Net Mayhill & New Mexico,USA \\
    & T11$^{\dagger}$ 50cm+FLI ProLine PL11002M & iTelescope.Net Mayhill & New Mexico, USA \\
CRI & 38cm K-380+Apogee E47 & Crimean astrophysical observatory & Crimea \\
deM & 29cm SC+QSI-516wsg & Observatorio Astronomico del CIECEM & Huelva, Spain \\
DPV & 28cmSC+MII G2-1600 & Astronomical Obs. on Kolonica Saddle & Slovakia \\
    & 35cmSC+MII G2-1600 & Astronomical Obs. on Kolonica Saddle & Slovakia \\
    & VNT 1m+FLI PL1001E & Astronomical Obs. on Kolonica Saddle & Slovakia \\
GFB & CDK 50cm+Apogee U6 & William Goff & California, USA \\
Ioh & 30cmSC+ST-9XE CCD & Hiroshi Itoh & Tokyo, Japan \\
Kai & 28cm SC+ST7XME & Kiyoshi Kasai & Switzerland \\
Kis & 25cm SC+Alta F47 & Seiichiro Kiyota & Kamagaya, Japan \\
KU2 & 40cm SC+Alta U6 & Kyoto U. Team & Kyoto, Japan \\
Mdy & 35cm SC+ST10XME & Yutaka Maeda & Nagasaki, Japan \\
OKU & 51cm+Andor DW936N-BV & OKU Astronomical Observatory & Osaka, Japan \\
Sac & 20cmL+ST-7XMEi & Atsushi Miyashita & Tokyo, Japan \\
SWI & C14 35cmSC+ST10XME & William L. Stein & New Mexico, USA \\
Trt & 25cm ALCCD5.2 (QHY6) & Tam\'{a}s Tordai & Budapest, Hungary \\
PXR & FTN 2.0m+E2V 42-40 & LCOGT$^{\ddagger}$ & Hawaii, USA \\
    & 35cmSC+SXV-H9 CCD & Roger D. Pickard & UK \\
\hline
\multicolumn{4}{l}{\parbox{420pt}{$^{*}$Observer's code: COO (Lewis M.~Cook), CRI (Cremean Observatory), deM (Enrique de Miguel), DPV (Pavol A.~Dubovsky), GFB (William Goff), Ioh (Hiroshi Itoh), Kai (Kiyoshi Kasai), Kis (Seiichiro Kiyota), KU2 (Kyoto Univ.~team), Mdy (Yutaka Maeda), OKU (Osaka Kyoiku Univ.~team), Sac (Atsushi Miyashita), SWI (William L.~Stein), Trt (Tam\'{a}s Tordai), PXR (Roger D.~Pickard).}}\\
\multicolumn{4}{l}{$^{\dagger}$itelescope.net.}\\
\multicolumn{4}{l}{$^{\ddagger}$Las Cumbres Observatory Global Telescope Network.}
	\end{tabular}
\end{table*}

\newpage

\begin{table*}
	\centering
	\caption{Log of observations of the outburst of V404 Cyg 
	from December, 2015 to January, 2016.}
	\label{log}
	\begin{tabular}{rrrrrccr}
		\hline
		${\rm Start}^{*}$ & ${\rm End}^{*}$ & ${\rm Mag}^{\dagger}$ 
		& ${\rm Error}^{\ddagger}$ & $N^{\S}$ & ${\rm Obs}^{\parallel}$ & 
		${\rm Band}^{\#}$ & exp[s]$^{\P}$\\
		\hline
80.5857 & 80.6428 & 16.471 & 0.010 & 67 & GFB & $CV$ & 60 \\
81.5472 & 81.5759 & 14.871 & 0.029 & 17 & Kis & $I_{\rm C}$ & 60 \\
81.5840 & 81.6288 & 14.908 & 0.055 & 19 & COO & $I_{\rm C}$ & 120 \\
82.1913 & 82.3047 & 14.762 & 0.016 & 97 & Kai & $I_{\rm C}$ & 30 \\
82.2616 & 82.3117 & 16.270 & 0.015 & 58 & deM & $CV$ & 60 \\
82.5703 & 82.5799 & 14.720 & 0.038 & 5 & FJQ & $I_{\rm C}$ & 60 \\
82.8506 & 82.8902 & 14.677 & 0.033 & 72 & Sac & $I_{\rm C}$ & 90 \\
82.8654 & 82.9298 & 14.469 & 0.016 & 76 & Ioh & $I_{\rm C}$ & 60 \\
83.2243 & 83.3247 & 14.641 & 0.017 & 96 & Kai & $I_{\rm C}$ & 30 \\
83.8717 & 83.9765 & 14.666 & 0.018 & 113 & Ioh & $I_{\rm C}$ & 60\\
83.9423 & 83.9762 & 2.092 & 0.023 & 54 & KU2 & $CV$ & 30 \\
84.1944 & 84.3028 & 14.438 & 0.029 & 121 & Kai & $I_{\rm C}$ & 30\\
84.8845 & 84.9610 & 1.778 & 0.008 & 194 & KU2 & $CV$ & 30 \\
84.8949 & 84.9487 & 14.536 & 0.016 & 58 & Ioh & $I_{\rm C}$ & 60\\
85.1498 & 85.2349 & 17.141 & 0.101 & 14 & CRI & $V$ & 540 \\
85.1520 & 85.2370 & 15.580 & 0.091 & 14 & CRI & $R_{\rm C}$ & 540 \\
85.1541 & 85.2327 & 14.455 & 0.082 & 13 & CRI & $I_{\rm C}$ & 540 \\
85.8582 & 85.9476 & 1.474 & 0.013 & 225 & KU2 & $CV$ & 30\\
85.8590 & 85.8949 & 14.469 & 0.036 & 46 & Kis & $I_{\rm C}$ & 60 \\
85.8742 & 85.9006 & 14.204 & 0.029 & 45 & Sac & $I_{\rm C}$ & 90 \\
85.8807 & 85.9853 & 14.177 & 0.027 & 61 & Ioh & $I_{\rm C}$ & 60 \\
85.8896 & 85.9168 & 14.440 & 0.026 & 25 & OKU & $I_{\rm C}$ & 90 \\
86.1662 & 86.2920 & 13.832 & 0.032 & 170 & DPV & $I_{\rm C}$ & 60 \\
86.2803 & 86.3182 & 14.721 & 0.017 & 44 & deM & $I_{\rm C}$ & 60 \\
86.5761 & 86.5853 & 14.761 & 0.022 & 8 & FJQ & $I_{\rm C}$ & 60 \\
86.5864 & 86.6248 & 15.484 & 0.017 & 48 & GFB & $R_{\rm C}$ & 60 \\
86.8507 & 86.8922 & 14.088 & 0.064 & 87 & Kis & $I_{\rm C}$ & 60 \\
86.8620 & 86.9545 & 14.479 & 0.032 & 155 & OKU & $I_{\rm C}$ & 90 \\
87.1656 & 87.2502 & 15.685 & 0.042 & 165 & Trt & $CV$ & 30 \\
87.1920 & 87.2546 & 13.993 & 0.084 & 52 & Kai & $I_{\rm C}$ & 30 \\
87.5755 & 87.6155 & 16.053 & 0.013 & 50 & GFB & $R_{\rm C}$ & 60 \\
87.8738 & 87.9768 & 13.161 & 0.027 & 141 & Ioh & $I_{\rm C}$ & 60 \\
87.8743 & 87.9645 & 0.332 & 0.023 & 206 & KU2 & $CV$ & 30 \\
88.1771 & 88.2466 & 14.972 & 0.075 & 97 & Trt & $V$ & 30 \\
88.2894 & 88.3206 & 13.840 & 0.058 & 44 & PXR & $I_{\rm C}$ & 60 \\
88.5759 & 88.6195 & 14.942 & 0.045 & 50 & GFB & $R_{\rm C}$ & 60 \\
88.8485 & 88.8857 & 13.585 & 0.055 & 70 & Kis & $I_{\rm C}$ & 60 \\
88.8652 & 88.9784 & 13.321 & 0.024 & 229 & Ioh & $I_{\rm C}$ & 60\\
89.5859 & 89.6227 & 16.246 & 0.011 & 46 & GFB & $R_{\rm C}$ & 60 \\
89.8472 & 89.8826 & 14.856 & 0.032 & 46 & Kis & $I_{\rm C}$ & 60 \\
89.9381 & 89.9754 & 13.794 & 0.020 & 79 & Ioh & $I_{\rm C}$ & 60 \\
89.9390 & 89.9591 & 0.822 & 0.031 & 42 & KU2 & $CV$ & 30 \\
90.8639 & 90.9495 & 14.549 & 0.020 & 60 & Ioh & $I_{\rm C}$ & 60 \\
90.8702 & 90.9571 & 1.711 & 0.022 & 115 & KU2 & $CV$ & 30 \\
90.8989 & 90.9337 & 15.957 & 0.026 & 45 & Mdy & $R_{\rm C}$ & 30 \\
91.1727 & 91.2256 & 17.334 & 0.049 & 75 & Trt & $CV$ & 60 \\
91.5676 & 91.6260 & 12.413 & 0.079 & 154 & SWI & $I_{\rm C}$ & 15 \\
91.8560 & 91.9038 & 15.126 & 0.044 & 40 & Sac & $I_{\rm C}$ & 90 \\
94.9056 & 94.9321 & 2.204 & 0.031 & 22 & KU2 & $CV$ & 30 \\
95.2544 & 95.2997 & 14.948 & 0.019 & 24 & PXR & $I_{\rm C}$ & 60\\
97.8620 & 97.8926 & 14.909 & 0.028 & 28 & Ioh & $I_{\rm C}$ & 60 \\
\hline
\multicolumn{8}{l}{$^{*}{\rm BJD}-2457300.0$.}\\
\multicolumn{8}{l}{$^{\dagger}$Mean magnitude.}\\
\multicolumn{8}{l}{$^{\ddagger}1\sigma$ of mean magnitude.}\\
\multicolumn{8}{l}{$^{\S}$Number of observations.}\\
\multicolumn{8}{l}{\parbox{260pt}{$^{\parallel}$\textcolor{black}{see the annotation in Table \ref{telescope} and FJQ (Foster James)}}}\\
\multicolumn{8}{l}{\parbox{260pt}{$^{\#}$Filter.  ``$I_{\rm C}$'', ``$R_{\rm C}$'', ``$V$'' and ``$CV$'' mean $I_{\rm C}$, $R_{\rm C}$, $V$ and no (clear) filter.}}\\
	\end{tabular}
\end{table*}

\newpage

\begin{table*}
	\centering
	\caption{Comparison stars in the photometric campaign of the 2015 December outburst in V404 Cyg.}
	\label{comparison-star}
	\begin{tabular}{cccc}
		\hline
		Observer$^{*}$ & Comparison star & RA (J2000) & Dec (J2000)\\
		\hline
COO & AUID 000-BCL-455 & 20:23:53.43 & $+$33:52:24.6\\
CRI & USNO1238.0435621 & 20:24:07.16 & $+$33:50:51.8\\
deM & AUID 000-BCL-468 & 20:24:08.89 & $+$33:54:38.6\\
DPV & UCAC4 620-101865 & 20:24:07.18 & $+$33:50:51.7\\
GFB & AUID 000-BCL-475 & 20:24:26.08 & $+$33:49:51.4\\
Ioh, Kai & AUID 000-BCL-460 & 20:23:56.47 & $+$33:48:16.9\\
Mdy, Kis, Sac & AUID 000-BCL-467 & 20:24:07.24 & $+$33:50:52.2\\
KU2 & USNO1200.15285418 & 20:24:18.72 & $+$33:53:13.9\\
OKU, Sac & AUID 000-BCL-476 & 20:24:28.31 & $+$33:51:13.5\\
SWI & AUID 000-BCL-471 & 20:24:59.41 & $+$33:57:56.6\\
Trt & AUID 000-BCL-472 & 20:24:18.66 & $+$33:53:12.6\\
PXR & AUID 000-BCL-458 & 20:23:55.26 & $+$33:51:14.8\\
\hline
\multicolumn{4}{l}{$^{*}$see the annotation in Table \ref{telescope}.}\\
\end{tabular}
\end{table*}

\newpage

\begin{figure}
\begin{center}
\includegraphics[width=7cm]{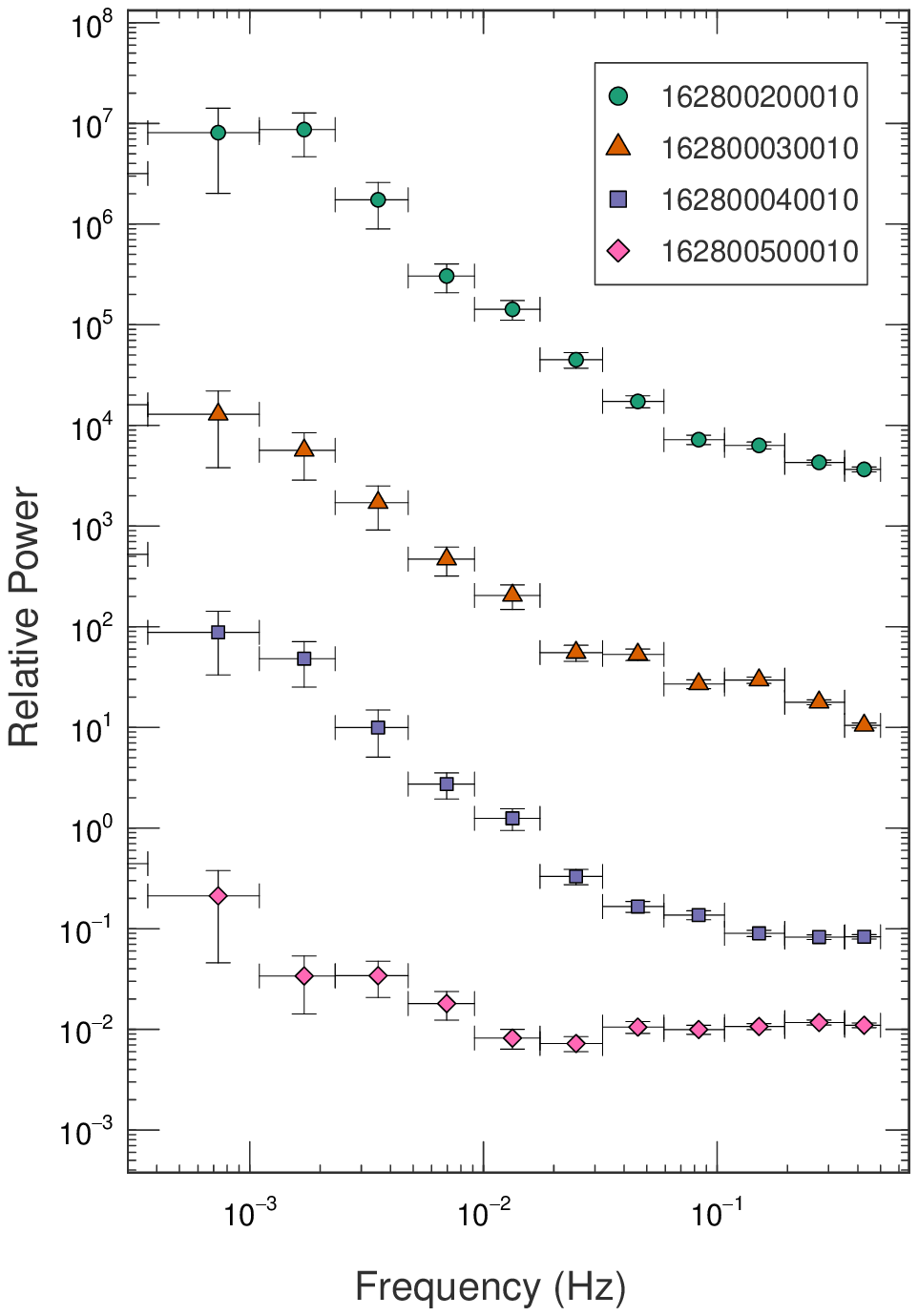}
\end{center}
\caption{PSDs of the X-ray light curves in the SCWs including 
interval (1) during the day 8.18--8.22 and interval (2) during the day 
8.246--8.292.  
The horizontal and vertical axes represent frequency and 
power in logarithmic scales.  For visibility, the powers in 
162800020010, 162800030010, 162800040010 and 162800050010 
are offset vertically by the value of $10^{3}$, $0$, 
$10^{-2}$ and $10^{-5}$, respectively.  The errors of PSDs 
represent $\pm$1$\sigma$.  }
\label{psd}
\end{figure}

\begin{figure}
\begin{center}
\includegraphics[width=8cm]{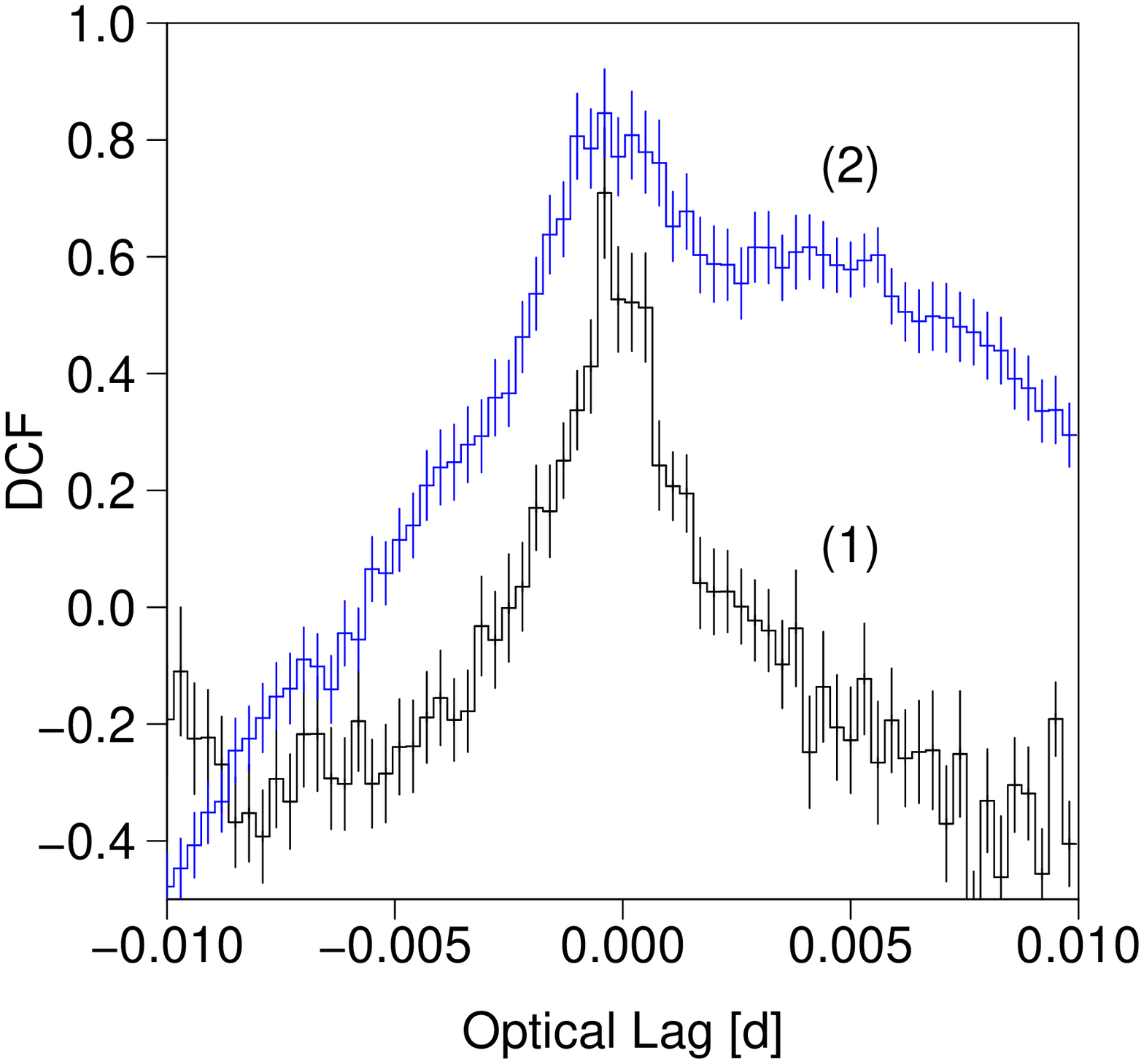}
\end{center}
\caption{Estimated LNDCFs in interval (1) during the day 8.18--8.22 
(black lines) and interval (2) during the day 8.246--8.292 (blue lines).  
The minus and plus signs in the horizontal axis indicate 
that X-ray emission is delayed to optical emission and that optical 
emission is delayed to X-ray emission.  We can see correlated 
positive peaks at the $-$35$_{-13}^{+13}$ s 
optical lags in the DCFs as for both two intervals.  }
\label{dcf}
\end{figure}



\newpage

\appendix

\section{Details of the Bayesian model, implementation, and model checking}

   We use a Bayesian model \citep{tak16timedelay} to estimate 
the time lag between optical and X-ray light curves.  The 
notation $\boldsymbol{x}=(x_1, \ldots, x_n)$ indicates the 
observed magnitudes of the X-ray light curve whose reported 
standard deviations of the measurement errors are 
$\boldsymbol{\delta}=(\delta_1, \ldots, \delta_n)$ at $n$ 
observation times $\boldsymbol{t_x}=(t_{x_1}, \ldots, t_{x_n})$.  
Similarly, the observed magnitudes of the optical light curve 
are $\boldsymbol{y}=(y_1, \ldots, y_m)$ with reported standard 
deviations of the measurement errors $\boldsymbol{\eta} = 
(\eta_1, \ldots, \eta_m)$ at $m$ observation times 
$\boldsymbol{t_y} = (t_{y_1}, \ldots, t_{y_m})$.  The number of 
observations for the X-ray light curve $n$ can be different from 
that for the optical light curve $m$ and the observation times 
of the X-ray light curve $\boldsymbol{t_x}$ can be different from 
those of the optical light curve $\boldsymbol{t_y}$.  
The model assumes that the observed magnitudes are generated from 
Gaussian distributions centred at the latent magnitudes with 
standard deviations of the measurement errors, i.e., 
\begin{align}
x_i \mid X(t_{x_i}) &\sim \textrm{N}\!\left(X(t_{x_i}),~\delta_i^2\right)~\textrm{for}~i=1, 2, \ldots, n,\label{obs_x}\\
y_j \mid Y(t_{y_j}) &\sim \textrm{N}\!\left(Y(t_{y_j}),~\eta_j^2\right)~\textrm{for}~j=1, 2, \ldots, m,\label{obs_y}
\end{align}
where $X(t_{x_i})$ and $Y(t_{y_j})$ are the latent magnitudes of 
the X-ray and optical light curves at times $t_{x_i}$ and $t_{y_j}$, 
respectively. A curve-shifted model \citep{pel94timedelay} assumes 
that the latent optical light curve is a shifted version of the 
latent X-ray light curve, i.e.,
\begin{equation}
Y(t_{y_j}) = X(t_{y_j} - \Delta)+\beta_0,\label{curve_shift}
\end{equation}
where $\Delta$ is the time delay in days and $\beta_0$ is the 
magnitude offset between the latent optical and X-ray light curves.  
Using \eqref{curve_shift}, we re-express \eqref{obs_y} as, for 
$j=1, 2, \ldots, m$,
\begin{equation}
y_j \mid X(t_{y_j}-\Delta), \Delta, \beta_0\sim \textrm{N}\!\left(X(t_{y_j} -\Delta)+\beta_0,~\eta_j^2\right)\!.\label{obs_y2}
\end{equation}

   We assume that the latent light curve follows 
a continuous-time damped random walk (DRW) process 
\citep{kel09QuasarOptical} whose stochastic differential equation 
is defined as
\begin{equation}
dX(t)=-\frac{1}{\tau}(X(t)-\mu)dt + \sigma dB(t),\nonumber
\end{equation}
where $\mu$ and $\sigma$ denote the overall mean and short-term 
variation of the DRW process on the magnitude scale, respectively, 
$\tau$ is a timescale of the process in days, and $B(t)$ is a 
standard Brownian motion.  The solution of this stochastic 
differential equation leads to the Gaussian distributions of 
the latent magnitudes as follows.  
We use the notation $\boldsymbol{t}^\Delta=(t^\Delta_1, \ldots, 
t^\Delta_{n+m})$ to denote the sorted vector of $n+m$ observation 
times among the $n$ observation times $\boldsymbol{t_x}$ and the 
$m$ time-delay-shifted observation times $\boldsymbol{t_y}-\Delta$. 
Then,
\begin{align} 
\begin{aligned}\label{ou_dist}
X(t_1^\Delta)&\sim \textrm{N}\!\left(\mu,~ \frac{\tau\sigma^2}{2}\right)\!, ~\textrm{and for}~ j=2, 3, \ldots, n+m,\\
~~X(t_j^\Delta)\mid X(t_{j-1}^\Delta) &\sim \textrm{N}\!\left(\mu + a_j\big(X(t_{j-1}^\Delta) - \mu\big) ,~\frac{\tau \sigma^2}{2}(1-a^2_j)\right)\!,
\end{aligned}
\end{align}
where $a_j =\exp(-(t_j^\Delta-t^\Delta_{j-1})/\tau)$ and we suppress 
conditioning on $\Delta$, $\mu$, $\sigma$, and $\tau$ in \eqref{ou_dist}.  

   Our independent and jointly proper prior distributions 
on the model parameters, i.e., $\Delta$, $\beta_0$, 
$\mu$, $\sigma^2$, and $\tau$, are 
\begin{align}
\begin{aligned}\label{prior_dist}
\Delta  & \sim\textrm{Uniform}(w_1,~ w_2)\\
\beta_0  & \sim\textrm{N}(0,~ 10^5)\\
\mu  & \sim\textrm{Uniform}(-30,~ 30)\\
\sigma^2  & \sim\textrm{inverse Gamma}(1,~ 1)\\
\tau^2  & \sim\textrm{inverse Gamma}(1,~ 2\times 10^{-7}),
\end{aligned}
\end{align}
where we fix the range of the Uniform distribution of $\Delta$ 
at $(-0.04, 0.04)$ for interval (1) and at 
$(-0.046, 0.046)$ for interval (2) to prevent 
spurious modes of the time delay on the margins of its space 
and to consider the longer observation period of 
interval (2).  
We choose the shape and scale parameters of the inverse Gamma 
distributions of $\tau$ and $\sigma^2$ in a way to constrain 
$\Delta$, considering shorter timescale of the observed data than 
gravitationally lensed light curves.  Further details of the 
motivation for the choice of prior distributions are given in 
Sections 2.5 of \citet{tak16timedelay}.  

   Our full posterior density, 
$\pi(\boldsymbol{X}(\boldsymbol{t}^{\Delta}), \Delta, \beta_0, 
\mu, \sigma^2, \tau\mid\boldsymbol{x}, \boldsymbol{y})$, is 
proportional to the multiplication of the probability density 
functions, whose distributions are specified in \eqref{obs_x}, 
\eqref{obs_y2}, \eqref{ou_dist}, and \eqref{prior_dist}.  
We use a Metropolis-Hastings within Gibbs sampler 
\citep{tie94markov} to sample the full posterior distribution 
of the model parameters; see Section 3 of \citet{tak16timedelay} 
for details of this sampler.  
To improve the convergence of the MCMC for $\Delta$ in the 
presence of multi-modality, we adopt a repelling-attracting 
Metropolis algorithm \citep{tak16multi}.  

   Before implementing the Bayesian method, we first checked 
the multi-modal behavior of $\Delta$ by quickly mapping a wide 
range of $\Delta$ between $-$86.40 s and 86.40 s using the 
profile likelihood function of $\Delta$ defined as 
\begin{equation}
L_{\textrm{prof}}(\Delta)=\max_{\beta_0, \mu, \sigma^2, \tau}L(\Delta, \beta_0, \mu, \sigma^2, \tau).  \nonumber
\end{equation}
This profile likelihood is proven to be a simple approximation 
to its marginal posterior distribution; see Section 4 of 
\citet{tak16timedelay}.  We confirmed that the highest 
mode is near $-$45 s and three weak modes are near $-$50 s and 
$-$30 s, and $-$25 s; the relative heights (the ratio of the 
profile likelihoods) of the modes near $-$50 s, $-$30 s and 
$-$25 s compared to the mode near $-$45 s 
are 5.7$\times$10$^{-3}$, 
5.1$\times$10$^{-8}$ and 6.5$\times$10$^{-9}$, respectively.  
In interval (2), the dominant mode is at around 
$-$33 s and an invisibly small mode is near $-$48 s; the relative 
height of the mode near $-$48 s compared to that of the mode near 
$-$33 s is 2.7$\times$10$^{-7}$.  

   We initialize three Markov chains near the dominating mode 
for each time interval, running for 150,000 
iterations; we discard the first 50,000 as burn-in iterations.  
The proposal scale of $\Delta$ (delta.proposal.scale in Table 
\ref{bayesian-parameter}) is set to produce the largest 
acceptance rate while making the Markov chains jump frequently 
between the modes identified by the profile likelihood.  The 
average acceptance rate of the time lag is 0.216 for 
interval (1) and 0.186 for 
interval (2).  

\begin{figure*}
\begin{minipage}{.49\textwidth}
\label{latent1}
\begin{center}
\includegraphics[width=8cm]{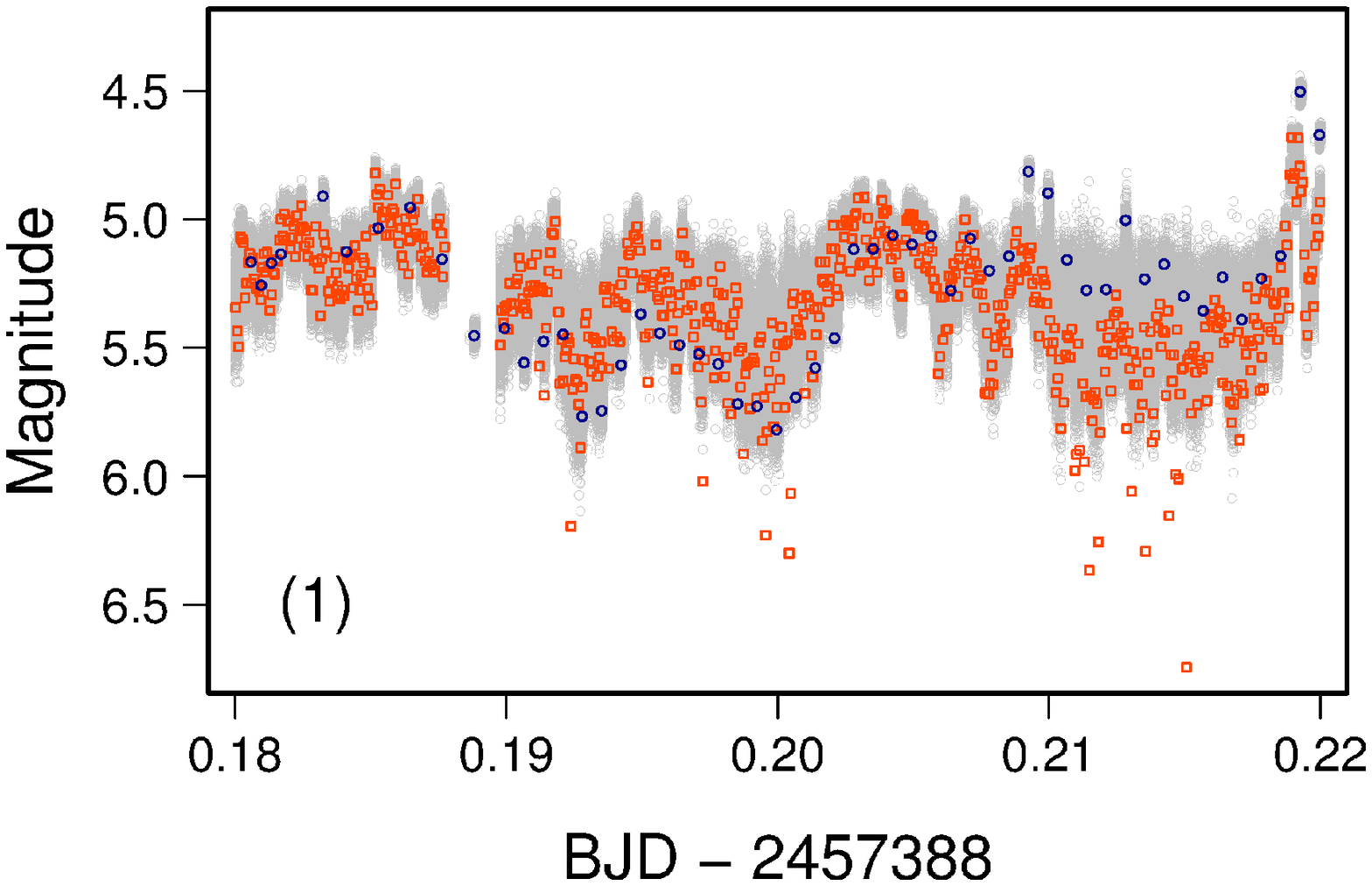}
\end{center}
\end{minipage}
\begin{minipage}{.49\textwidth}
\label{latent2}
\begin{center}
\includegraphics[width=8cm]{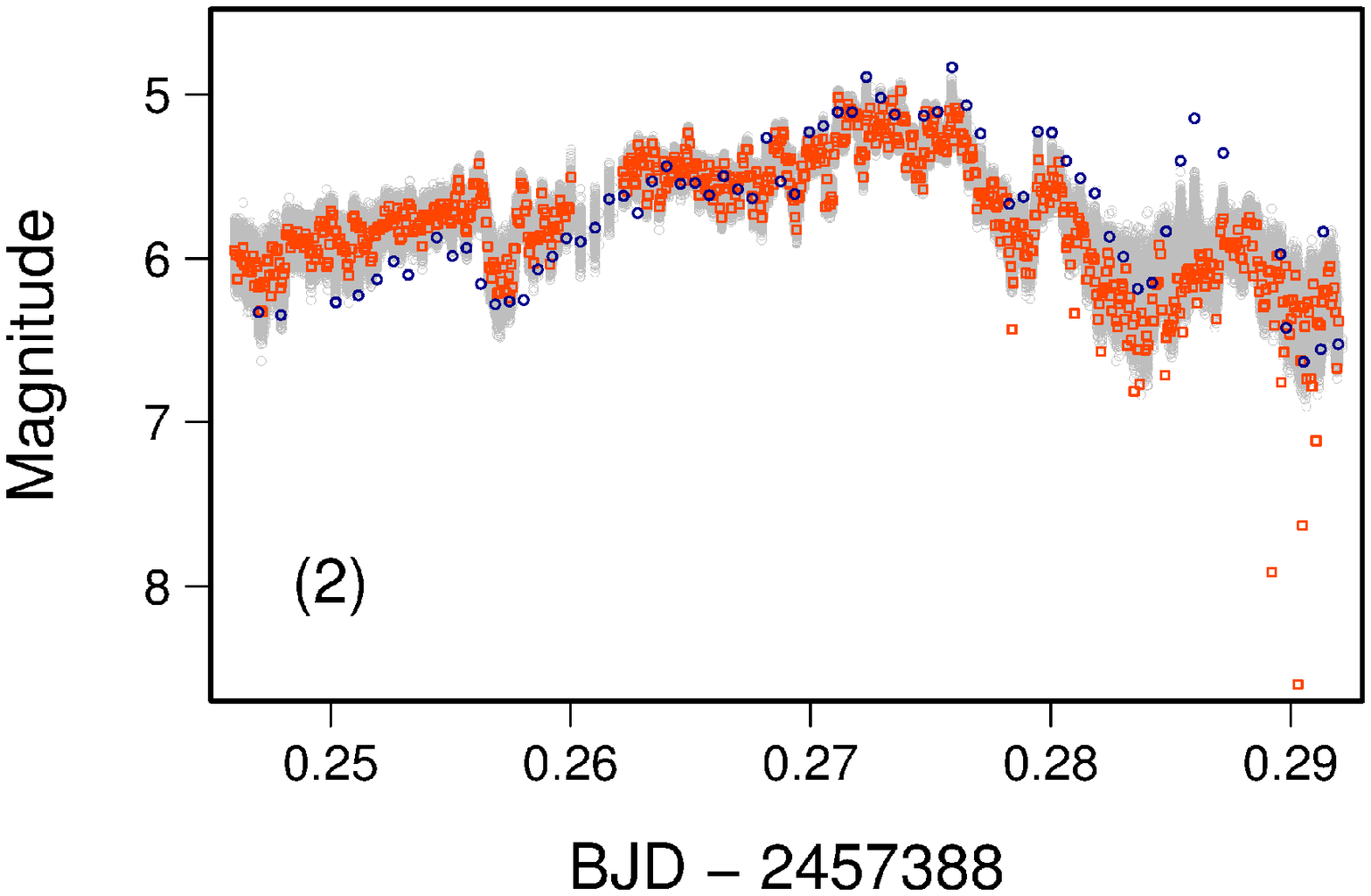}
\end{center}
\end{minipage}
\caption{The X-ray light curves are denoted by orange squares, the optical light curves are denoted by blue circles, and the posterior samples of latent light curves at intervals of 300 iterations are denoted by grey circles in interval (1) on the day 8.18--8.22 (the left panel) and interval (2) on the day 8.246--8.292 (the right panel).  Each optical light curve is shifted by the posterior mode of the time lag in the horizontal axis and by that of the magnitude offset in the vertical axis. The fitted model makes a good match of the fluctuations of the two light curves.  The grey areas are encompassing most of the light curves, meaning that the fitted model describes the observed data well.  }
\label{latentLC}
\end{figure*}

   We visually checked our estimates, shifting the optical light 
curves by the posterior medians of $\Delta$ and $\beta_0$ in 
Figure \ref{latentLC}; the fitted model matches the fluctuations 
of the two light curves well.  We also conducted a model checking 
by plotting the posterior sample of the latent curve in grey; the 
grey areas encompass most of the observed light curves, which shows 
how well the fitted model predicts the observed data.  
A sensitivity analysis, though not shown here, exhibits that our 
inferential results in Table \ref{bayesian-result} are robust to 
changing the scale parameters in the inverse Gamma distributions of 
$\sigma^2$ and $\tau$.  

\begin{table}
	\centering
	\caption{Initial values of the parameters for each of three 
	Markov chains used in the function \texttt{bayesian} of the 
	R package \texttt{timedelay} for interval 
	(1) during the day 8.18-8.22 and interval 
	(2) during the day 8.246-8.292.  }
	\label{bayesian-parameter}
	\begin{tabular}{ccc}
		\hline
		Names of parameters & (1) & (2)\\
		\hline
theta.ini ($\mu$, $\sigma$, $\tau$)$^{*}$ & (5.30, 100, 0.01) & (5.72, 100, 0.01)\\
                                          & (5.30, 10, 0.1) & (5.72, 10, 0.1) \\
                                          & (5.30, 1, 1) & (5.72, 1, 1) \\
delta.ini$^{\dagger}$ & $-$0.0005827546 & $-$0.0004405093\\
                      & $-$0.0005248843 & $-$0.0003826389\\
                      & $-$0.0004670139 & $-$0.0003247685\\
delta.uniform.range$^{\ddagger}$ & ($-$0.04, 0.04) & ($-$0.046, 0.046)\\
delta.proposal.scale$^{\S}$ & 0.00005 & 0.00005\\
tau.proposal.scale$^{\#}$ & 1 & 1\\
tau.prior.shape$^{\P}$ & 1 & 1\\
tau.prior.scale$^{**}$ & 2/10$^{7}$ & $2/10^{7}$\\
sigma.prior.shape$^{\dagger\dagger}$ & 1 & 1\\
sigma.prior.scale$^{\ddagger\ddagger}$ & 1 & 1\\
adaptive.delta$^{\S\S}$ & FALSE & FALSE\\
multimodality$^{\#\#}$ & TRUE & TRUE\\
micro$^{\P\P}$ & 0 & 0\\
\hline
\multicolumn{3}{l}{\parbox{220pt}{$^{*}$Initial values of the DRW parameters.  Unit of magnitudes in $\mu$ and $\sigma$.  Unit of days in $\tau$.}}\\
\multicolumn{3}{l}{\parbox{220pt}{$^{\dagger}$Initial value of the delay time for MCMC used in three Markov chains.  Unit of days.}}\\
\multicolumn{3}{l}{\parbox{220pt}{$^{\ddagger}$Range of uniform prior distribution of the time delay $\Delta$.  Unit of days.}}\\
\multicolumn{3}{l}{\parbox{220pt}{$^{\S}$Proposal scale of the Metropolis step for the time delay $\Delta$.  Unit of days.}}\\
\multicolumn{3}{l}{\parbox{220pt}{$^{\#}$Proposal scale of the Metropolis-Hastings step for $\log(\tau)$.  Units of $\tau$ are days.}}\\
\multicolumn{3}{l}{\parbox{220pt}{$^{\P}$Shape parameter of Inverse-Gamma hyper-prior distribution for $\tau$.  }}\\
\multicolumn{3}{l}{\parbox{220pt}{$^{**}$Scale parameter of Inverse-Gamma hyper-prior distribution for $\tau$.  }}\\
\multicolumn{3}{l}{\parbox{220pt}{$^{\dagger\dagger}$Shape parameter of Inverse-Gamma hyper-prior distribution for $\sigma^{2}$.  }}\\
\multicolumn{3}{l}{\parbox{220pt}{$^{\ddagger\ddagger}$Scale parameter of Inverse-Gamma hyper-prior distribution for $\sigma^{2}$.  }}\\
\multicolumn{3}{l}{\parbox{220pt}{$^{\S\S}$We do not use the adaptive MCMC for the time delay $\Delta$ in the presence of multi-modality because the adaptation may occur at a local mode.  }}\\
\multicolumn{3}{l}{\parbox{220pt}{$^{\#\#}$We use a repelling-attracting Metropolis algorithm to sample the time delay $\Delta$.  }}\\
\multicolumn{3}{l}{\parbox{220pt}{$^{\P\P}$The order of a polynomial regression model.  We do not consider the effect of microlensing in the case of V404 Cyg.  }}\\
\end{tabular}
\end{table}

\bsp	
\label{lastpage}
\end{document}